\documentclass[aps,pre,preprint,groupedaddress,showpacs]{revtex4-1}
\usepackage[dvips]{graphicx}
\usepackage{textcomp}
\usepackage{amssymb}

\newcommand{\mc}{\multicolumn}
\begin{document}

\title{
\Large\bf Fighting topological freezing in the two-dimensional CP$^{N-1}$ model
}

\author{Martin Hasenbusch}
\email[]{Martin.Hasenbusch@physik.hu-berlin.de}
\affiliation{
Institut f\"ur Physik, Humboldt-Universit\"at zu Berlin,
Newtonstr. 15, 12489 Berlin, Germany}

\date{\today}

\begin{abstract}
We perform Monte Carlo simulations of the CP$^{N-1}$ model on the square 
lattice for $N=10$, $21$, and $41$. Our focus is on the severe slowing
down related to instantons. To fight this problem we employ
open boundary conditions as proposed by L\"uscher and Schaefer for
lattice QCD. Furthermore we test the efficiency 
of parallel tempering in a line defect. Our results for open boundary 
conditions are consistent with the expectation that topological
freezing is avoided, while autocorrelation times are still large. 
The results obtained with parallel tempering are encouraging. 
\end{abstract}

\maketitle

\section{Introduction}
The CP$^{N-1}$ model shares certain fundamental properties such as asymptotic 
freedom and confinement with QCD.
Therefore this model has been frequently studied as a toy model of QCD. 
It has been shown \cite{Adda78,Witten79} that the model has a non-trivial vacuum 
structure with stable instanton solutions. It turned out that these
topological objects pose a particular problem in the simulation of the 
lattice CP$^{N-1}$ model, similar to lattice QCD. 

On the torus, in the continuum limit, the configuration space is decomposed into sectors
that are characterized by their topological charge. At finite lattice spacing, the 
free energy barriers between such sectors increase as the lattice spacing decreases.
For Markov chain Monte Carlo algorithms that walk  in a quasi continuous fashion through 
configuration space this means that they become essentially non-ergodic and slowing down 
becomes dramatic. Numerical results are compatible with an increase of autocorrelation 
times that is exponential in the inverse lattice spacing. 
In the case of the CP$^{N-1}$ model this was first suggested and numerically 
verified in ref. \cite{campostrini}. 
Later this behaviour was numerically confirmed for example in refs. 
\cite{Vicari04,Flynnetal15}. Modelling the autocorrelation times with a more conventional 
power law Ansatz, large powers are needed to fit the data.
From a practical point of view, the consequence is that it becomes virtually impossible
to access lattice spacings below a certain threshold. The numerical studies show that in the case of the 
CP$^{N-1}$ model the problem becomes worse with increasing $N$. 
Since it is much less expensive to simulate the two-dimensional model than lattice QCD, 
it is a good test bed for new ideas and algorithms that could overcome the severe 
slowing down of the topological modes. For example simulated tempering \cite{MaPa92} has been studied
in ref. \cite{Vicari92} with moderate success. More recently, 
``trivializing maps in the Hybrid Monte Carlo algorithm'' \cite{Engel11} 
or the ``Metadynamics'' method \cite{Metadynamics} have been tested. 

A very principle solution of the problem had been suggested in ref. \cite{LuSc11}. By abandoning
periodic boundary conditions in one of the directions in favour of open ones, barriers between 
the topological sectors are abolished. This idea should work independently  of 
the algorithm that is used in the simulation. The proposal has been further tested
\cite{LuSc13,McGlMa14} and adopted in large scale simulations of lattice QCD with 
dynamical fermions, see for example refs. \cite{ALPHA14,CLS}.

Here we shall probe in detail how open boundary conditions  effect the slowing 
down  in the case of the CP$^{N-1}$ model. Since the CP$^{N-1}$ model is much cheaper
to simulate than lattice QCD, a larger range of lattice spacings can be studied and 
autocorrelation functions can be computed more accurately.

Furthermore, we shall explore parallel tempering \cite{raex,SW86} 
as a solution to our problem. Parallel tempering is a well 
established approach in statistical
physics to overcome effective non-ergodicity due to a ragged free energy landscape.  
The idea of parallel tempering  and similar methods
is to enlarge the configuration space such that the hills can be easily 
by-passed. 
A prototype problem is the study of spin-glasses, where parallel tempering 
is mandatory. For recent work see for example ref. \cite{Janus2013}.
Typically a global parameter such as the temperature or
an external field is used as parameter of the tempering. Here instead, we shall discuss 
a localized defect. In particular we shall interpolate between a system with a line 
defect and a homogeneous system with periodic boundary conditions.

Finally we like to mention that for the CP$^{N-1}$ model dual formulations can be found. These can be 
simulated by using the worm algorithm \cite{wurm,Ulli}. In these dual formulations there are no topological
sectors and hence severe slowing down does not occur in the simulation.
For recent work and a discussion of the literature see ref. \cite{RiFo16}.  Unfortunately, 
a similar approach to full QCD has not been worked out yet.

The outline of the paper is the following: In the next section we define the lattice
model and the observables that we measure. Next we discuss the 
basic update algorithms and the parallel tempering scheme that are used. 
We perform standard simulations of lattices with periodic boundary conditions 
in both directions. We check that our results are consistent with those 
presented in the literature.
Then follow our simulations with open boundary conditions in one of the 
directions. Next we discuss our runs with parallel tempering.
We compare our results for physical quantities with those of the large 
$N$-expansion. Finally we conclude and give an outlook.

\section{The model}
We simulate the CP$^{N-1}$ model on the square lattice. It is defined 
by the action
\begin{equation}
\label{action}
 S = - \beta N \sum_{x,\mu}  \left(\bar{z}_{x+\hat \mu} z_x \lambda_{x,\mu} + z_{x+\hat \mu} \bar{z}_x
 \bar{\lambda}_{x,\mu} -2  \right) \;,
\end{equation}
where $z_x$ is a complex $N$-component vector with $z_x \bar{z}_x = 1$  and $\lambda_{x,\mu}$ is 
a complex number with $\lambda_{x,\mu} \bar{\lambda}_{x,\mu} = 1$. The sites of the lattice are
denoted by $x=(x_0,x_1)$, where $x_i \in \{0,1,2,...,L_i-1\}$. 
The lattice spacing is set to $a=1$ in the following. This means that we 
trade a decreasing
lattice spacing for an increasing correlation length. The gauge fields
live on the links, which are denoted by $x, \mu$, where $\mu \in \{0,1\}$ gives the direction.
$\hat \mu$ is a unit vector in $\mu$-direction.  In $1$-direction we shall always employ periodic
boundary conditions. In $0$-direction either periodic or open boundary conditions are employed.
Here we implement open boundary conditions in a crude way, simply setting $\beta=0$ for the links 
that connect $x_0=L_0-1$ and $x_0=0$.

\subsection{The observables}
\label{Sobservables}
In the case of periodic boundary conditions in both directions we 
follow the literature. For completeness we recall the definitions of
the observables that we measure.

The energy $E$ is given by 
\begin{equation}
E = \frac{1}{2 L_0 L_1} \sum_{x,\mu}
\langle 2- \bar{z}_{x+\hat \mu} z_x \lambda_{x,\mu} - z_{x+\hat \mu} \bar{z}_x \bar{\lambda}_{x,\mu} \rangle \; .
\end{equation}

Further observables are based on the composite operator
\begin{equation}
 P_x = \bar{z}_x \otimes z_x
\end{equation}
or in terms of the components
\begin{equation}
 P_{x,ij}  = \bar{z}_{x,i} z_{x,j} \;.
\end{equation}
The connected correlation function is now defined as 
\begin{equation}
 G_P(x) = \mbox{Tr} \langle P(x) P(0) \rangle - \frac{1}{N} \;.
\end{equation}
The Fourier transform of the correlation function is given by 
\begin{equation}
\tilde G(k) = \frac{1}{V} \sum_{x,y} \left[\mbox{Tr} \langle P(x) P(y) \rangle - \frac{1}{N} \right]
\exp\left(i \sum_{j=0}^{d-1} \frac{2 \pi}{L_j} (x_j-y_j) k_j \right) \;\;.
\end{equation}

The magnetic susceptibility is given by
\begin{equation}
\label{chidef}
 \chi=\sum_x G_P(x) = \tilde G(0) = \frac{1}{V} \sum_{x,y} \left[\mbox{Tr} \langle P(x) P(y) \rangle - \frac{1}{N} \right]
=  \frac{1}{V}   \mbox{Tr} \left \langle \left(\sum_{x} P(x) \right)^2 \right \rangle - \frac{V}{N}  \; .
\end{equation}

The second moment correlation length is defined as 
\begin{equation}
\label{xi2nd} 
\xi_{2nd}^2 = \frac{\mu_2}{2 d \chi} = \frac{\sum_x x^2 G_P(x)}{2 d \sum_x  G_P(x)}  \;, 
\end{equation}
where $d$ is the dimension of the system. In Monte Carlo simulations 
this definition has to be adjusted to the finite size of the lattice.
A popular choice for the second moment correlation length on a finite lattice is
\begin{equation}
\label{xi2murks}
\xi_{2nd}^2 = \frac{1}{4 \sin^2 (\pi/L)} \left(\frac{\tilde G(0,0)}{\tilde G(1,0)} - 1 \right) \;. 
\end{equation}
A downside of this definition is that it introduces $O(L^{-2})$ corrections 
that are avoided by other definitions; see for example 
eqs.~(\ref{chidefopen},\ref{mu2defopen}) below.

A geometrical definition of the topological charge is given by 
\cite{BergLuescher81}  
\begin{equation}
\label{Qtri} 
Q_{tri} = \frac{1}{2 \pi} \sum_x \mbox{Im} \left(\log \mbox{Tr} [P_{x+\hat \mu + \hat \nu} P_{x+\hat \mu} P_x ] 
+ \log \mbox{Tr} [P_{x+ \hat \nu} P_{x+\hat \mu +\hat \nu} P_x ] \right) \;,
\end{equation}
where the principle branch of the logarithm is taken and $\mu \ne \nu$.  
Motivated by eq.~(33) of ref. \cite{campostrini}  we have used
\begin{equation}
\label{Qplaq}
Q_{plaq} = \frac{1}{2 \pi} \sum_x  \theta_{plaq,x}  \;,
\end{equation}
where 
\begin{equation}
\theta_{plaq,x} =  \theta_{x,\mu} +  \theta_{x + \hat \mu,\nu} 
 -\theta_{x + \hat \nu,\mu} - \theta_{x,\nu} - 2 n \pi \;\;\;,\;\;\;
\;\;\; \mu \ne \nu \;,
\end{equation}
where $\theta_{x,\mu}= \mbox{arg}\{\bar{z}_x z_{x+\hat \mu}\}$
and the integer $n$ is chosen such that $-\pi < \theta_{plaq,x} \le \pi$. 
The topological susceptibility is then given by
\begin{equation}
\label{chitper}
\chi_t = \frac{1}{V} \langle Q^2 \rangle \; .
\end{equation}
Note that the definitions~(\ref{Qtri}) and (\ref{Qplaq}) are not equivalent 
at finite lattice spacing. For $N=10$ at $\beta=0.8$, which is the smallest 
value of $\beta$ that we have studied, we find that eq.~(\ref{Qplaq}) gives
a value for the topological susceptibility that is about $3 \%$ smaller than
that obtained by using  eq.~(\ref{Qtri}).  This difference decreases
with increasing $\beta$. The difference is roughly proportional to $\xi^{-2}$. 
For larger values of $N$, the difference seems to decrease even more 
rapidly. We also did a few experiments with cooling. After one step of 
our cooling procedure, the difference in the topological susceptibility 
obtained by using the two different definitions of the topological charge
is drastically reduced. We conclude that both definitions lead to the same 
results in the continuum limit. In most of our simulations, we only determined
the topological susceptibility obtained with eq.~(\ref{Qplaq}), since it 
requires somewhat less CPU time than  eq.~(\ref{Qtri}).  The numbers reported 
in the following always refer to  eq.~(\ref{Qplaq}). 

\subsubsection{Open boundary conditions in $0$-directions}
The definitions of susceptibilities and the second moment correlation length
given above have to be adjusted to open boundary conditions. This has
been discussed for example in section 3.2 of ref. \cite{ALPHA14} for the 
topological susceptibility in the case of lattice QCD. 

Let us start with the definition of the magnetic susceptibility. 
First we rewrite the definition~(\ref{chidef}) for periodic boundary conditions
such that it is suitable for generalization to open ones:
\begin{eqnarray}
\label{chidef2}
 \chi&=&\frac{1}{L_0 L_1} 
 \mbox{Tr} \left \langle \left(\sum_{x} P(x) \right)^2 \right \rangle -
    \frac{L_0 L_1}{N}  
=\frac{1}{L_0 L_1}
 \mbox{Tr} \left \langle \left(\sum_{x_0} \sum_{x_1}  P(x) \right)^2 \right \rangle -
    \frac{L_0 L_1}{N}  \nonumber
\\
&\simeq& 
 \frac{1}{L_0}  \sum_{x_0} \sum_{t=-t_{max}}^{t_{max}}
  \left[ \mbox{Tr}  \langle  S_{x_0} S_{x_0+t} \rangle -\frac{L_1}{N} \right]
=  \frac{1}{L_0} \sum_{x_0} \left[G_S(x_0,0) + 2 \sum_{t=1}^{t_{max}} G_S(x_0,t) \right]
\;,
\end{eqnarray}
where
\begin{equation} 
S_{x_0} = \frac{1}{\sqrt{L_1}} \sum_{x_1} P(x) \;.
\end{equation}
We assume that $G_S(x_0,t) = \mbox{Tr} \langle S_{x_0}  S_{x_0+t} \rangle -\frac{L_1}{N}$ 
decays as $\exp(-|t|/\xi_{exp})$ for large $|t|$ and $L_0 \gg t_{max} \gg \xi_{exp}$.
In order to avoid effects due to open boundaries we restrict the 
summation over $x_0$ such that we stay away from the boundaries by the 
distance $l_0$:
\begin{equation}
\label{chidefopen}
 \chi_{open}= \frac{1}{L_0-2 l_0} \sum_{x_0=l_0}^{L_0-l_0-1} G_S(x_0,0)
            + 2 \sum_{t=1}^{t_{max}}
\frac{1}{L_0-2 l_0-t}
  \sum_{x_0=l_0}^{L_0-l_0-t-1}  G_S(x_0,t) \;,
\end{equation}
where $t_{max} \gg \xi_{exp}$. 
The effects of the open boundary conditions decay exponentially fast with 
increasing $l_0$, where we assume $L_0 \gg l_0$. The exponential decay
is governed by the correlation length. Therefore $l_0$ should 
be chosen as a multiple of the correlation length $ \xi_{exp}$. Below
in section \ref{opensimul} we give a quantitative discussion of this point.
In order to get the 
second moment correlation length~(\ref{xi2nd}) we compute
\begin{equation}
\label{mu2defopen}
\mu_{2,open} =
 2 \sum_{t=1}^{t_{max}}
\frac{1}{L_0-2 l_0-t}
  \sum_{x_0=l_0}^{L_0-l_0-t-1} t^2 \; G_S(x_0,t) \;.
\end{equation}
The topological susceptibility is computed in an analogous fashion, taking 
into account the topological charge in the interior of the lattice.
Details
of the implementation are discussed below in section \ref{opensimul}.

\section{The algorithms}
In this section we discuss the update schemes that we use in our 
simulations. For our preliminary simulations with periodic boundary conditions 
in both directions
and for the simulations with open boundary conditions in $0$-direction we use
a hybrid of different local updates. In order to reduce the autocorrelation 
times of the topological susceptibility in the case of periodic boundary 
conditions in both directions, we employ a parallel tempering 
scheme \cite{raex,SW86}. In the last part of this section we recall
the definitions of the integrated and the exponential autocorrelation
time and discuss how they can be determined from the data.

\subsection{Basic local algorithms}
As basic algorithm  we use a hybrid of the Metropolis, the heatbath
and the microcanonical overrelaxation algorithm.  To a large extend, we follow section III of ref. \cite{campostrini}.
Let us first discuss the updates of the site variables and then the update of the gaugefields.

In an elementary step of the algorithm  we update the variable at
a single site $x$, while keeping the gauge fields and the variables at all other 
sites fixed. The part of the action that depends on this site variable can be written as
as
\begin{equation}
 \tilde S(z_x) = - \mbox{Re} \; z_x \bar{F}_x \;,
\end{equation}
where 
\begin{equation}
 F_x = 2 N \beta \; \sum_{\mu}  
   \left [\bar{\lambda}_{x,\mu} z_{x+\hat \mu} +  
          \lambda_{x-\hat \mu,\mu} z_{x-\hat \mu}  \right] \;.
\end{equation}
Note that the problem at this point is identical to the update of an
$O(2 N)$ invariant vector model with site variables of unit length. Instead 
of $\bar{F}_x$ we would have to deal with the sum of the variables on the nearest neighbour 
sites. The microcanonical update keeps $\tilde S(z_x)$ fixed, while the new 
value of $z_x$ has maximal distance from the old one. It is given by eq.~(43a)
of ref. \cite{campostrini}:
\begin{equation}
 z_x' = 2 \frac{ \mbox{Re} \; z_x \bar{F}_x}{|F_x|^2} F_x  - z_x \;.
\end{equation}
In addition to these updates, we have to perform updates that change the value
of the action. To this end we implemented a heat-bath algorithm that is applied 
to the subset of three of the $2 N$ components of $z_x$. The heatbath used 
here is identical to the one used in the simulation of the O(3)-Heisenberg model
on the lattice or for the update of $SU(2)$ subgroups in the simulation of
pure $SU(N)$ lattice gauge models \cite{CM82,Cr80}. 
We run through all $N$ components 
of $z_x$ taking the real and the complex part of the component as first two components for
the heat-bath. The third component is randomly chosen among the real or complex parts of the
remaining $N-1$ components of $z_x$.  
Note that the CPU-time required by the microcanonical overrelaxation update is about
one order of magnitude less than that for the heat-bath update.

For fixed variables $z$ the gaugefields can be updated independently of each other. The action
reads
\begin{equation}
 \tilde S_g(\lambda_{x,\mu}) = - \mbox{Re} \lambda_{x,\mu} \bar{f}_{x,\mu} \;,
\end{equation} 
where 
\begin{equation}
 f_{x,\mu} =  2 N \beta \; z_{x+\hat \mu} \bar{z}_x \;.
\end{equation}
Here we performed a four hit Metropolis update, where the stepsize was chosen such that the
acceptance rate is roughly $50\%$, 
and a microcanonical update 
\begin{equation}
\lambda_{x,\mu}' =  2 \frac{ \mbox{Re} \; \lambda_{x,\mu} \bar{f}_{x,\mu}}{|f_{x,\mu}|^2} f_{x,\mu}  - \lambda_{x,\mu}      \;\;, 
\end{equation}
see eq.~(43b)  of \cite{campostrini}.  

In our standard simulations of lattices with periodic boundary conditions, 
we  organized the elementary updates of site and link variables in 
the following way: First we sweep through the lattice in lexicographic
order, updating the site variables by using the heat-bath algorithm. Then we update all
link variables using the four hit Metropolis update.  Next we sweep 
$n_{ov}$ times through the lattice by using the microcanonical overrelaxation 
algorithm. Also here we go through the lattice in lexicographic order. 
For each sweep over the sites, microcanonical overrelaxation updates 
of the gaugefields are performed. In our simulations, we have chosen
$n_{ov} \propto \xi$, which is for example the recommended
choice for the simulation of the two-dimensional XY-model \cite{Guptaetal}.
The choices made in the design of our update scheme are based on a few
preliminary simulations but still some ad hoc decisions are taken.

In the case of our simulations with open boundary conditions, 
we were aiming at the simulation of rather large correlation lengths
and lattice sizes. Therefore, we parallelized the simulation 
program using the Message Passing Interface (MPI).  For 
simplicity,  we split the lattice in $0$-direction only.
In order to simplify the parallelization, we used a checker-board 
decomposition of the lattice. First
the variables on the odd sites are updated and than the ones on the
even sites. Since odd sites have only even neighbours and vice
versa, the updates on odd (even) only, can be done in parallel.
Preliminary tests  indicate that the ordering, lexicographical
versus checker-board decomposed, has little influence on the 
autocorrelation times. 

A measurement of the observables is  performed after a heat-bath 
sweep and $n_{ov}$ overrelaxation sweeps are completed.

\subsection{Parallel tempering in a line defect}
\label{ParallelT}
If one insists on specific boundary conditions in temporal direction that are
different from open ones, parallel tempering or related methods 
might be an option to avoid the severe slowing down.
Note that in  ref. \cite{Vicari92},  simulated tempering with
$\beta$ as parameter had been used in the simulation of the CP$^{N-1}$ model;
according to ref. \cite{Vicari04}
``..., but apparently without achieving a particular advantage.'' Since 
simulated tempering is closely related with parallel tempering we did not 
investigate parallel tempering in $\beta$  here.
Instead, motivated by the nice results 
obtained by the simulations with open boundary conditions, see section 
\ref{opensimul} below, we consider parallel tempering restricted to a defect 
line. Let us sketch the idea: 
We expect that in the system with the defect line, 
the topological charge of a configuration is quickly changed by 
local updates.
Then, via tempering, the configuration moves to the homogeneous system 
that we are interested in. The faster the configurations move from the 
bottom to the top of the parallel tempering chain, the more efficient is the
algorithm.

In a preliminary study
we introduced a sequence of systems that interpolate between open and periodic boundary conditions.
Later we used instead a line defect, where the linear extension $l_d$ of the defect is smaller than
the linear lattice size $L_1$. The advantage is that less interpolating systems are needed
for $l_d<L_1$ than for open boundary conditions. On the other hand, the smaller $l_d$, the 
less topological objects can be created or destroyed in a unit of time. Below in section
\ref{NumTempering} we  determine the optimal value of $l_d$ numerically.
In order to define the interpolating systems, we generalize the action~(\ref{action}) to
\begin{equation}
\label{actionT}
 S = - \beta N \sum_{x,\mu} \;  c_{t,x,\mu}  \; \left(\bar{z}_{t,x+\hat \mu} z_{t,x} \lambda_{t,x,\mu} + z_{t,x+\hat \mu} \bar{z}_{t,x}
 \bar{\lambda}_{t,x,\mu} -2     \right) \;,
\end{equation}
where $c_{t,x,\mu}=c_r(t)$  for $x_0=L_0-1$, $x_1<l_d$ and $\mu=0$ and $c_{t,x,\mu}=1$ else.
In the case of open  boundary conditions $l_d=L_1$.
$t \in \{0,1,...,N_t-1\}$ labels the points of the tempering chain.
In our simulations we take for simplicity a linear interpolation:
$c_r(t)=1-t/(N_t-1)$. The homogeneous system corresponds to $t=0$, while for 
$t=N_t-1$ the coupling along the defect line is completely eliminated. Since
we have a configuration for each $t$, we add an index $t$ to the field variables.
In order to perform a swap of the configurations,
only the contribution of the action at the defect $x_0=L_0-1$, $x_1<l_d$, $\mu=0$ has to be computed. To this
end we define
\begin{equation}
 E_r(t) = - \sum_{x_1=0}^{l_d-1} \;  \left(  \bar{z}_{t,(0,x_1)} z_{t,(L_0-1,x_1)} \lambda_{t,(L_0-1,x_1),0}
                            + z_{t,(0,x_1)} \bar{z}_{t,(L_0-1,x_1)}  \bar{\lambda}_{t,(L_0-1,x_1),0} -2
 \right) \;.
\end{equation}
A swap of configurations between $t_1$ and $t_2$  is accepted with
the probability
\begin{equation}
\label{Aswap}
 A_{swap} = \mbox{min} \left[1,\exp\left(-\beta N \left[c_r(t_1) - c_r(t_2)\right]
 \left[E_r(t_2) - E_r(t_1)\right]\right) \right] \;.
\end{equation}
In our simulations we run from $t_1=0$ up to $t_1=N_t-2$ in steps of one, 
proposing to swap the configurations at $t_1$ and $t_2=t_1+1$. The number of
replica $N_t$ is chosen such that the acceptance rate for the swap of 
configurations is larger than $30 \%$ for all $t_1$. 
In a tempering simulation, updates of the individual configurations, using for example the local
Metropolis algorithm, alternate with swaps of the configurations. In a simulation
with a global tempering parameter such as the temperature, one usually performs  
sweeps over the whole lattice in between swaps. In our case the tempering part of the
action is localized. Furthermore, the correlation of $E_r$ with the field decays with the
distance from the defect line.
Therefore one might not necessarily always perform updates of all field variables between
the swaps of configurations, allowing for a larger number of swaps for a given 
amount of CPU-time. To this end,
we performed updates of the field variables on the sites and the gaugefields for a rectangle that
is centred around the defect line. The linear extension of the rectangle is  $\mbox{min}[L_0,2 l_i]$
in 0-direction and $\mbox{min}[L_1,l_d+2 l_i]$ in $1$-direction.  The index $i$ of $l_i$ indicates
the level of our hierarchical update scheme. With increasing level $i$, the size of the rectangle is reduced.
Let us explain this update scheme by using a piece of pseudo-C code, representing a single cycle of the update scheme:
\begin{verbatim}

Sweeps over the full lattice;   replica exchange;  translation;
  for(i1=0;i1<n1;i1++)
    {
    Sweeps over box(l_1);  replica exchange;  translation;  measure;
    for(i2=0;i2<n2;i2++)
      {
      Sweeps over box(l_2);  replica exchange;  translation;
      for(i3=0;i3<n3;i3++)
        {
        Sweep over box(l_3);  replica exchange;  translation;
        .
        .
        .  until l_i = 1
    } } ...}

\end{verbatim}

\begin{figure}
\begin{center}
\includegraphics[width=15.0cm]{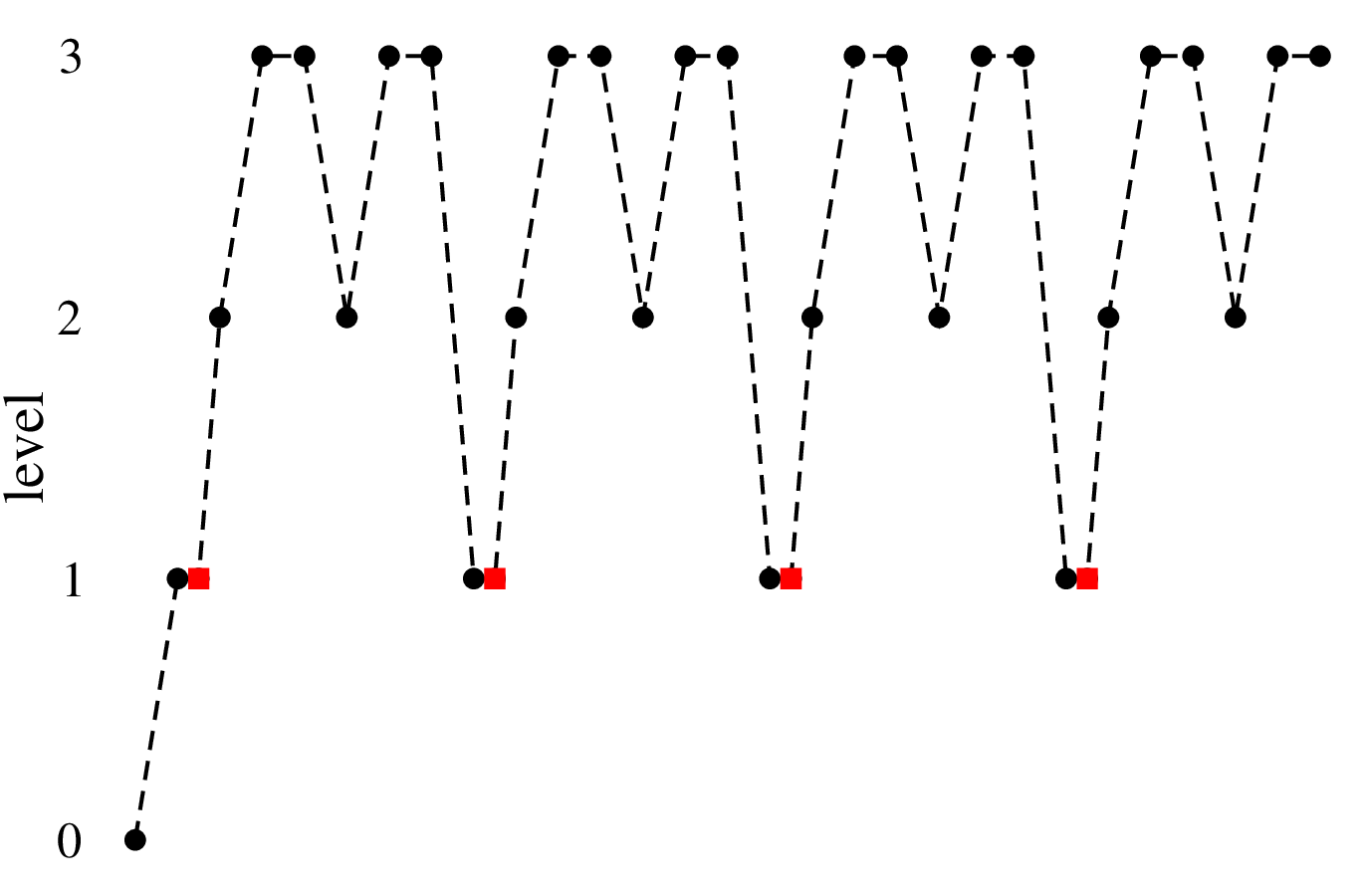}
\caption{\label{levelplot}
We give a graphical representation of one cycle of our parallel 
tempering scheme. Time evolves from left to right.  
The number of levels and the values
$n_1=4$, $n_2=n_3=2$ are selected so that the plot is readable. In our
simulations we use a larger number of levels and $n_1$ is typically larger.
A circle stands for a sequence of sweeps over the rectangle associated 
with the given level for all $t$, a swap-update as discussed in the text and 
a translation of the configuration at $t=0$. 
Costs are dominated by the sweeps over the rectangles. Note that the 
size of the rectangle decreases with increasing level. 
At level $0$ we sweep over the  whole lattice.
The squares indicate the measurements. Measurements are performed
for the homogeneous system only and involve all sites of the lattice.
Note that the figure gives correctly the order of the steps, 
the CPU-time required by the single steps is however not proportional 
to the interval taken on the $x$-axis. The dashed lines should only guide
the eye. For a detailed discussion see the text.
}
\end{center}
\end{figure}

In addition, in Fig. \ref{levelplot} we give a graphical
representation of one update cycle.
Let us go through the code step by step. ``\verb|Sweeps over the full lattice|'' means that we
perform one sweep over the whole lattice by using the heatbath update for the variables
on the sites and the Metropolis update of the gaugefield. Then follow $n_{ov}$ sweeps
using the overrelaxation update of the variables on the sites as well as the gaugefields.
``\verb|Sweeps over box(l_i)|'' means that we sweep over a rectangle characterized by $l_i$.
In particular for small values of $l_i$ we use a number of overrelaxation updates that
is smaller than the one for sweeps over the whole lattice. In our numerical experiments
we use $l_{i+1} =l_i/2$ throughout. $l_1$ is taken as an integer power of $2$ and
roughly $l_1\approx \xi$. ``\verb|replica exchange|'' is the swap of 
configurations as discussed below eq.~(\ref{Aswap}).
An important ingredient of the update is the  ``\verb|translation|''.
Note that for $c_r=1$ we have restored the translational invariance of the system. Therefore
we can shift the configuration for $t=0$ by a random distance in both directions.  The idea
is that topological objects  are created or destroyed at any location in the lattice and
need not to diffuse.
Note that actively performing the translation would be computationally too expensive. Therefore
we actually randomly chose a new location of the defect line for the system with $c_r=1$.
Performing swaps, we have to keep track of this location. 
 In a preliminary study we switched
off the translations. It turned out that the performance of the algorithm degrades markedly.

We performed no fine tuning of the
parameters $n_1$, $n_2$, $n_3$ ... .  Instead we tried to balance the CPU-time needed for the
different levels of the update scheme.  For $n_{ov}$ being the same for all levels $i$,
this  means
\begin{equation}
 n_i \approx  \frac{2 l_{i-1} (l_d + 2 l_{i-1})}{2 l_i (l_d + 2 l_i)} \;\;.
\end{equation}

``\verb|measure|'' means a calculation of the observables discussed in
section \ref{Sobservables}.  The measurement is only performed for the
system $t=0$, $c_r=1$.  In principle one would like to measure after each
swap of configurations, since new configurations are shuffled to $t=0$.
However, since the measurement of the observables involve the field on the
whole lattice, it costs more CPU-time than the updates at high levels
of the update scheme. As a compromise, we perform the measurements
along the update at level $1$ of our scheme. In our program we average
over the $n_1$ measurements performed within one complete update cycle.
These averages are written to the data file.
In the analysis of our data, such a complete cycle is the unit of time
in our Markov chain. 

In order to compare with the standard update of the
system with periodic boundary conditions, we have to take into account
the additional effort.
Ignoring that $n_{ov}$ is not the same for all levels we arrive at the factor
\begin{equation}
N_t \times N_{level}  \;,
\end{equation}
where $N_t$ is the number of replica and $N_{level}$ the number of levels
of the update scheme.

We parallelized our program using the Message Passing Interface (MPI).
To this end we distributed the configurations among the processes. As usual
in parallel tempering simulations, we do not copy the configurations to
new locations in memory. Instead the value of $t$ that is associated with
the configurations is swapped. 

\subsection{Autocorrelation times} 
The performance of a Markov chain Monte Carlo algorithm is characterized
by the autocorrelation time. There are different definitions of the 
autocorrelation time. These are based on the autocorrelation function.
The autocorrelation function of an estimator $A$ is given by
\begin{equation}
 \rho_A(t) = \frac{\langle A_{i} A_{i+t} \rangle - \langle  A \rangle^2}
                  {\langle A^2 \rangle - \langle  A \rangle^2} \;.
\end{equation}
The modulus of the autocorrelation function is bounded from above 
by an exponentially decaying function. Following ref. \cite{Sokal} we define
\begin{equation}
\tau_{exp,A} = \lim_{t\rightarrow \infty}  \mbox{sup} \frac{t}{-\log(|\rho_A(t)|)}
\end{equation}
and
\begin{equation}
\tau_{exp}=  \mbox{sup}_A  \tau_{exp,A} \;,
\end{equation}
which characterizes the Markov chain. If the transition probabilities
of the Markov chain satisfy detailed balance, the autocorrelation function
is given by an exponential decay at large $t$. Therefore it is useful to 
compute the effective autocorrelation time 
\begin{equation}
\label{taueff}
 \tau_{eff,s,A}(t) = -\frac{s}{ \ln\left(\rho(t+s)/\rho(t)  \right) } \;\;,
\end{equation}
where $s=1$, $2$, ... is the stride. With increasing $t$ it approaches
$\tau_{exp,A}$. 
The integrated autocorrelation time of the estimator $A$ is given by
\begin{equation}
\label{tauint}
 \tau_{int,A} = 0.5 + \sum_{t=1}^{\infty} \rho_A(t) \;.
\end{equation}
The statistical error of the estimate of  $\langle A \rangle$ is
\begin{equation}
\epsilon_A = \sqrt{\frac{2 \tau_{int,A}}{N} \sigma_A^2} \;,
\end{equation}
where $N$ is the number of measurements and 
\begin{equation}
\sigma_A^2 = \langle A^2 \rangle - \langle  A \rangle^2
\end{equation}
is the variance. Computing $\tau_A$, 
we use the estimate of $\rho_A(t)$ obtained from our simulation. Therefore 
the summation in eq.~(\ref{tauint}) has to be truncated at some finite $t_{max}$. 
Since $\rho_A(t)$
is falling off exponentially at large distances, the relative statistical
becomes large at large distances. Therefore it is mandatory to truncate
the summation at some point that is typically much smaller than the 
total length of the simulation.  In the literature one can find various 
recommendations how this upper bound should be chosen. Sokal \cite{Sokal} 
suggest to take $t_{max} \approx M \tau_{int}$, 
where the value of $M$ should be at least $6$.  Wolff \cite{Wolffless} proposes to 
balance the statistical error with the systematic one that is due to the 
truncation of the sum. 

In both cases it is assumed that the ratio $\tau_{int,A}/\tau_{exp,A}$ is not
too far from one. In systems like the one discussed here, this assumption
is not satisfied for all observables. 
It turns out that $\tau_{exp}$ can be  associated with the 
topological modes. It can be cleanly observed in the topological 
susceptibility for example. However, it also enters into the autocorrelation
function of other quantities like the susceptibility with a
small amplitude. In such a situation, the rules of Sokal and Wolff are not
suitable. In ref. \cite{Virotta}  the approach of
ref. \cite{Wolffless} was extend to take such a slow mode with a small 
amplitude explicitly into account.


\section{Numerical results}
In this section we present our numerical results.  We study the 
performance of the algorithms, where we focus on the autocorrelation times 
of the topological susceptibility.  First we discuss our standard simulations 
of systems with periodic boundary. Then we report the results for systems
with open boundary conditions in $0$-direction. Finally we discuss our 
parallel tempering simulations of systems with periodic boundary conditions.
We implemented the code in standard C and used the SIMD-oriented Fast Mersenne Twister
algorithm \cite{twister} as random number generator.

\subsection{Standard simulation of periodic boundary conditions}
\label{standardperiodic}
To set the scene,
we performed preliminary simulations with periodic boundary conditions.
We started with $N=10$, where we compared with the extensive study 
presented in \cite{Flynnetal15}. Then we performed simulations for 
$N=21$ and $41$. Note that the slowing down of the topological charge
becomes more severe with increasing $N$, see for example refs. 
\cite{Vicari92,campostrini,Vicari04}. 

\subsubsection{N=10}
The values of $\beta$ and the lattice size $L$ were chosen to match a few of 
those of tables 6.2 and 6.3 of ref. \cite{Flynnetal15}.  Note that the authors 
of ref. \cite{Flynnetal15}
checked carefully for finite size effects. For the runs reported in  tables 6.2 and 6.3
they have used a linear lattice size $L \approx 15 \xi_{2nd}$.  Note that
the authors of ref. \cite{Flynnetal15} used the over-heatbath algorithm 
\cite{petronzio,campostrini}, which is 
similar but not identical to the hybrid of heatbath and overrelaxation used here.
We performed preliminary
simulations at $\beta=0.8$ with various values of $n_{ov}$. 
It turns out that the 
performance has a quite shallow dependence on $n_{ov}$. At the end we took
$n_{ov}=20$ for our more extend run. For other values of
$\beta$ we scaled $n_{ov}$ proportional to the correlation length.
Our numerical results are summarized in table \ref{Operiodic}. All 
simulations were started with an ordered configuration.
For $\beta \le 0.96$, we performed  $10^6$ times one heatbath update and $n_{ov}$ 
microcanonical overrelaxation updates along with the corresponding updates of the
gaugefields. In the case of $\beta =1.0$ only $9 \times 10^5$ such update cycles
were performed. With increasing autocorrelation time, we discarded an increasing 
number of measurements at the beginning of the simulations. In the case of
$\beta=1$, we discarded $10^5$ measurements.
Note that we performed a measurement after one sweep
with the heatbath update and $n_{ov}$ microcanonical overrelaxation updates.
The integrated autocorrelation times are given in units of these measurement.
While the  integrated autocorrelation time of the topological susceptibility increases
rapidly with increasing correlation length, the  integrated autocorrelation times
of the energy density and the magnetic susceptibility stay of order one.
However one should note that in particular for the magnetic susceptibility 
there is a small overlap with $\tau_{exp}$. This is most significant for 
our largest value of $\beta=1.0$. Using the program of ref. \cite{Virotta} 
we find $\tau_{int,\chi} = 1.15(14)$ instead of $0.754(5)$, truncated at 
$t_{max}=7$. 

\begin{figure}
\begin{center}
\includegraphics[width=14.5cm]{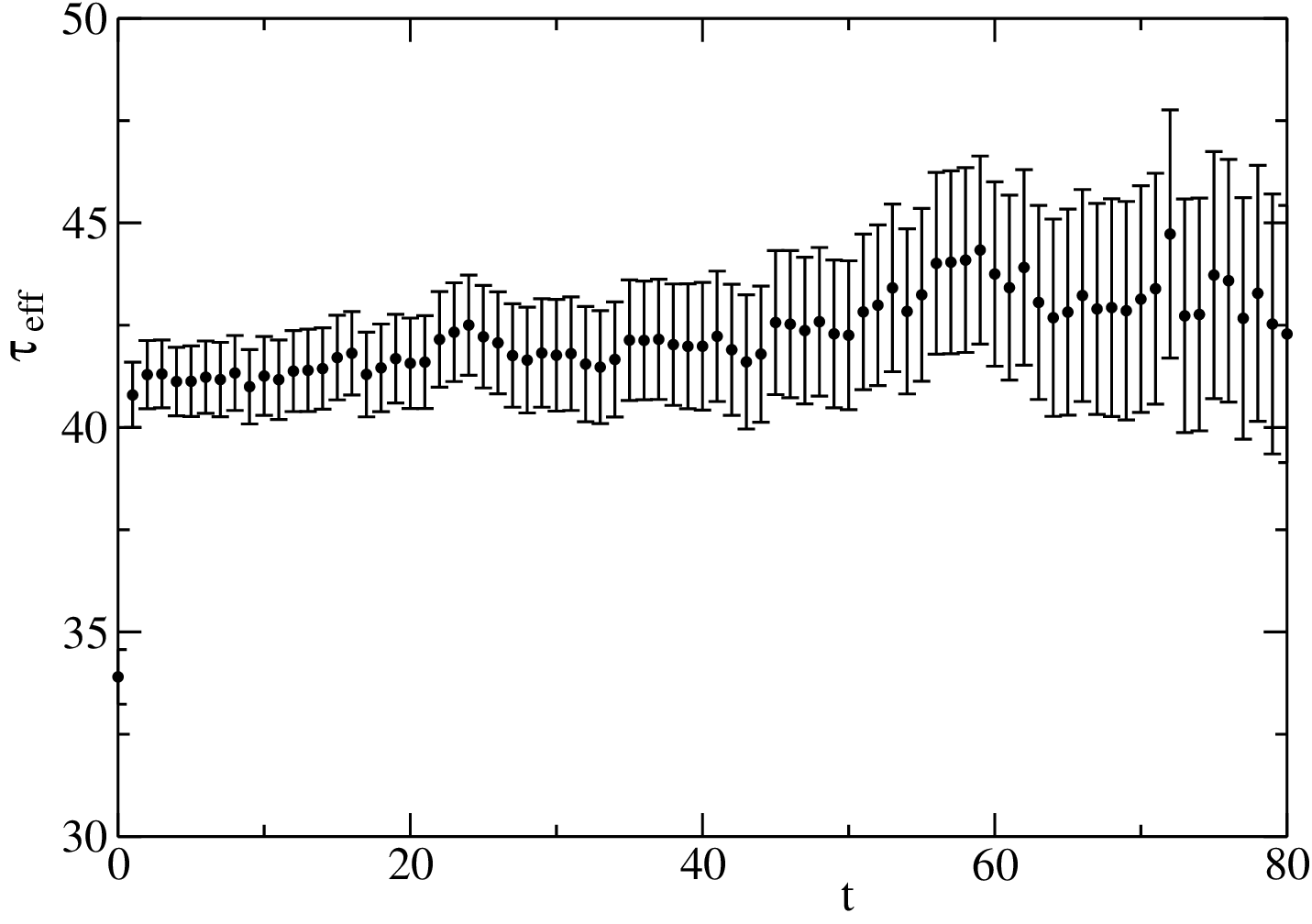}
\caption{\label{taueff96}  We plot the effective autocorrelation time of
the topological susceptibility
$\tau_{eff,1,\chi_t}$ as a function of the distance $t$ in the Markov chain. 
The data are taken from the standard simulation with periodic boundary 
conditions, $N=10$ at $\beta=0.96$.
}
\end{center}
\end{figure}

We tried to extract the exponential autocorrelation time $\tau_{exp}$ 
from the  autocorrelation function of the topological susceptibility. 
To this end we computed the effective autocorrelation time~(\ref{taueff}).
In Fig. \ref{taueff96}, as an example,
we plot $\tau_{eff,1,\chi_t}(t)$ for $\beta=0.96$. We see that 
$\tau_{eff,1,\chi_t}(t)$ rapidly approaches a plateau. The value that 
is approached at this plateau is consistent with $\tau_{int,\chi_t}$.
This means that the autocorrelation function of the topological susceptibility 
is given to a good approximation by a single exponential decay. 
This finding is consistent with the prediction based on the effective model
given in ref. \cite{McGlMa14}. We checked
that the same  holds for any of the values of $\beta$ studied here. 
Furthermore we find analogous results for $N=21$ and $41$. 

In order to compare with ref. \cite{Flynnetal15} we multiply our values for $\tau$ by $1+n_{ov}$.
In case of the topological susceptibility we find that our autocorrelation times are smaller by 
about a constant factor of $1.25$. This means that the scaling with the correlation 
length should be the same.  Note that in ref. \cite{Flynnetal15} results for 
coupling constants up to
$\beta=1.05$ are given, where the second moment correlation length is 
$\xi_{2nd}=22.558(13)$ and the integrated autocorrelation time of the 
topological susceptibility is $\tau_{int,\chi_t}=70900(4500)$.

\begin{table}
\caption{\sl \label{Operiodic}
Parameters and numerical results of our standard simulations with periodic boundary 
conditions and $N=10$. We give the value of $\beta$, the linear lattice size $L$ and 
the number of overrelaxation sweeps per update cycle $n_{ov}$. It follows the 
energy density $E$, the magnetic susceptibility $\chi$, the second moment correlation length
$\xi_{2nd}$, the topological susceptibility $\chi_t$ and finally, in the last column,
the integrated autocorrelation time $\tau_{int,\chi_t}$ of the topological susceptibility.
In order to compute $\tau_{int,\chi_t}$, we truncated the summation of the autocorrelation
function at $t_{max} \approx 6 \tau_{int,\chi_t}$. 
}
\begin{center}
\begin{tabular}{lccccccc}
\hline
\mc{1}{c}{$\beta$} & $L$ & $n_{ov}$ & $E$     & $\chi$     & $\xi_{2nd}$ & $\chi_t$ & $\tau_{int,\chi_t}$ \\
\hline
0.8 &  72 & 20 &  0.6670180(54) &  28.0588(53) & 4.5970(28) &  0.0009711(22) & 1.19(1)  \\
0.85&  96 & 28 &  0.6222722(36) &  46.8669(90) & 6.3965(37) &  0.0004651(17) & 3.02(5) \\
0.9 & 136 & 38 &  0.5838375(23) &  78.186(17)  & 8.8113(50) &  0.0002299(14) & 9.30(25) \\
0.96& 192 & 56 &  0.5440439(15) & 144.896(32)  & 12.8674(74)&  0.0001042(14) & 43.3(2.4)\\
1.00& 248 & 72 &  0.5205875(11) & 219.052(64)  & 16.524(11) &  0.0000616(15) & 131.(14.) \\
\hline
\end{tabular}
\end{center}
\end{table}

\subsubsection{N=21}
The authors of \cite{campostrini}  simulated the model using a hybrid
of local Metropolis and microcanonical overrelaxation updates. They simulated up to
$\beta=0.75$, however only quote results for $\chi_t$ up to $\beta=0.65$, where
$\xi_{2nd} \approx 2.7$. In ref. \cite{Vicari92} results up to $\beta=0.72$, where
$\xi_{2nd} =4.23(2)$, are given. Note that in ref. \cite{Vicari92} simulated
tempering in $\beta$ was used.
 In ref. \cite{Vicari04} the authors simulated the CP$^{N-1}$ model with a Symanzik
improved action by using a hybrid of  local Metropolis and microcanonical
overrelaxation updates.
For $N=21$ they went up to a correlation length $\xi_{2nd} \approx 4.2$,
where they find an integrated autocorrelation time of
$\tau_{int,\chi_t} = 19000(3000)$ for the topological susceptibility.
In the recent work \cite{Metadynamics}, the metadynamics method was tested.
As basic  update algorithm the Hybrid-Monte-Carlo (HMC) algorithm \cite{HMC}
was used. By using
the plain HMC algorithm, the authors were able to produce reliable estimates of the 
topological susceptibility up to $\beta=0.75$, where $\xi_{2nd} \approx 5.15$.
Employing the metadynamics method  they could extend the range of the correlation 
length up to $\xi  \approx 13.3$ at $\beta=0.9$, albeit the accuracy
of their estimate $\chi_t =0.000042(23)$ at $\beta=0.9$ is moderate.
For a discussion of the  metadynamics method  we refer the reader to 
ref. \cite{Metadynamics}.

Our results are summarized in table \ref{P21}. In our simulations for $N=21$ we used
a smaller ratio $n_{ov}/\xi_{2nd}$ as for $N=10$ to allow for more measurements. 
Throughout we used lattices of the linear size $L \gtrapprox 11 \xi_{2nd}$. 
Here still finite size effects can be seen at our level of statistical accuracy.
However, this should be sufficient to study the scaling of the autocorrelation 
time of the topological susceptibility with the correlation length. For 
$\beta=0.625$, $0.65$, and $0.675$ we performed $2 \times 10^6$ update cycles
and the corresponding measurements. For $\beta=0.7$, $0.725$, and $0.75$ we 
performed only $10^6$ measurements.  
The number of discarded configurations increases with $\beta$. 
In the case of $\beta=0.75$ we discarded the first $3 \times 10^5$ 
measurements. 

The integrated autocorrelation time $\tau_{int,\chi_t}$ increases rapidly with 
increasing correlation length. Fitting all data, we get 
$\tau_{int,\chi_t} = 0.275(28) \times \exp(1.72(4) \times \xi_{2nd})$ 
with $\chi^2/$d.o.f.$=0.89$.  Instead, fitting with the Ansatz
$\tau_{int,\chi_t} = a \exp(b \xi_{2nd}^{\theta})$  with $\theta=1/2$, 
as suggested in ref. \cite{Vicari04}, gives $\chi^2/$d.o.f.$=3.3$.

We also performed simulations for larger values of $\beta$.
For $\beta=0.8$ we get
a measurement of $\chi_t \ne 0$ for the first time after 159201 update cycles.
A sufficient equilibration is never reached in our run.
In the case of $\beta=0.9$, throughout our $10^6$ update cycles $\chi_t = 0$.

Our data are not good enough to decide on the functional form of the increase
of $\tau_{int,\chi_t}$ with increasing correlation length.
But it is clear that with standard algorithms and periodic boundary conditions
it is practically impossible to get reliable results  for the 
topological susceptibility for $\xi_{2nd} \gtrapprox 6$. 

\begin{table}
\caption{\sl \label{P21}
Parameters and numerical results of our standard simulations with periodic boundary
conditions and $N=21$. We give the value of $\beta$, the linear lattice size $L$ and
the number of overrelaxation sweeps per update cycle $n_{ov}$. It follows the
energy density $E$, the magnetic susceptibility $\chi$, the second moment correlation length
$\xi_{2nd}$, the topological susceptibility $\chi_t$ and finally, in the last column,
the integrated autocorrelation time $\tau_{int,\chi_t}$ of the topological susceptibility.
In order to compute $\tau_{int,\chi_t}$, we truncated the summation of the autocorrelation
function at $t_{max} \approx 6 \tau_{int,\chi_t}$.
}
\begin{center}
\begin{tabular}{cccccccc}
\hline
$\beta$ & $L$ & $n_{ov}$ & $E$     & $\chi$     & $\xi_{2nd}$ & $\chi_t$ & $\tau_{int,\chi_t}$ \\
\hline
0.625  &26 &  3 &  0.833348(10)& 9.648(2)    & 2.2879(6) &  0.001696(9) &    14.16(35) \\
0.65   &32 &  4 &  0.799334(8) & 12.179(2)   & 2.6993(7) &  0.001170(9) &    28.76(90) \\
0.675  &38 &  5 &  0.768003(6) & 15.429(3)   & 3.1814(8) &  0.000820(9) &    63.1(2.5) \\
0.7    &44 &  6 &  0.739044(7) & 19.603(6)   & 3.744(2)  &  0.000570(16) &   199.(18.) \\
0.725  &52 &  7 &  0.712213(7) & 24.954(11)  & 4.398(3)  &  0.000426(20) &   524.(64.) \\
0.75   &60 &  8 &  0.687255(5) & 31.86(3)    & 5.165(8)  &  0.00027(3)&  2600.(1000.)  \\
\hline
\end{tabular}
\end{center}
\end{table}

Later, as preliminary study for our simulations with open boundary conditions
and parallel tempering, we performed additional simulations at $\beta=0.675$ 
for the 
linear lattice sizes $L=32$, $44$, $52$, $60$ and $80$. From the analysis 
of these simulations we conclude that for $L/\xi_{2nd} \gtrapprox 16$ finite 
size effects can not be detected at our level of accuracy.

\subsubsection{N=41}
First we performed simulations at $\beta=0.55$ for $L=18$, $24$, $28$,
$32$, and $40$ to check for finite size effects. We took $n_{ov}=3$.
Throughout we performed $2 \times 10^6$ measurements.
In the case of the energy density and the susceptibility, we find that 
the results are consistent among each other within the statistical error
starting from $L=28$. In the case of $\xi_{2nd}$ we see differences up 
to our largest lattice size. This is likely due to the $O(L^{-2})$ corrections
that are intrinsic to the definition~(\ref{xi2murks}) of $\xi_{2nd}$. 
In the case 
of the topological susceptibility, we find that the results are consistent 
starting from $L=24$. The same holds for the integrated autocorrelation time
of the topological susceptibility. For $L=28$ we get $\tau_{int,\chi_t}=239.(36.)$. 
 Extrapolating $\xi_{2nd}$ in $L$, assuming 
 $O(L^{-2})$ corrections we arrive at $\xi_{2nd}=1.766(2)$ at  $\beta=0.55$. 
Assuming scaling, we conclude that, similar to $N=10$ and $21$, 
for $L/\xi_{2nd} \gtrapprox 16$ finite  size effects can be ignored at the
level of our statistical accuracy. 

Finally we performed  simulations at $\beta=0.62$, $0.65$, $0.7$, 
and $0.75$ with 
100000 measurements each. For $\beta=0.62$, the value of the topological 
susceptibility changed five times. For $\beta=0.65$, $0.7$, and $0.75$ 
it remained at zero throughout the whole simulation. We conclude that it is 
practically impossible to go beyond $\beta \approx 0.62$ simulating 
periodic boundary conditions using standard Monte Carlo algorithms.
For $\beta =0.62$ we find by interpolation of numerical results 
presented below $\xi_{2nd} \approx 2.6$.  Note that in ref. \cite{Vicari92} 
$\beta=0.6$ could be reached, where $\xi_{2nd} =2.431(11)$. 

\subsection{Open boundary conditions in $0$-direction}
\label{opensimul}
In our simulations, similar to lattice QCD,
open boundary conditions are introduced
to avoid the freezing of the topological charge in the simulation.
The effects of the open boundary conditions on the observables are
however unwanted. These effects decay exponentially fast  with the
distance from the boundaries. Therefore one discards the sites with a distance
smaller than $l_0$ in the measurements of the observables. The value
of $l_0$ is chosen such that the systematic error remains below a certain
threshold. These systematical errors should
be smaller than the statistical error that is reached in the simulations.
Since the fraction of the lattice that is discarded should not be too large,
one takes $L_0$ a few times $L_1$.  On the other hand,
the diffusion of instantons from the boundary
to the centre of the lattice requires more time with increasing $L_0$.
After a few preliminary simulations  we fixed $L_0 = 4 L_1$ for all three
values of $N$. Furthermore we checked that $l_0 \approx 10 \xi_{2nd}$
is sufficient to avoid systematic errors that are larger than our
statistical ones.
In our simulations, we write an estimate of $\chi_{open}$,
eq.~(\ref{chidefopen}), $\mu_{2,open}$, eq.~(\ref{mu2defopen}), and
$\chi_{t,open}$ for one value of $t_{max}$ for each measurement into the data file.
After a few preliminary simulations we decided $t_{max} = l_0$.
These numbers are used to compute  autocorrelation functions.
In addition, we wrote the estimates of $G_S(x_0,t)$ and
the corresponding correlation function of the topological charge for each
value $t \le t_{max}$ to a file. Here we first averaged over
$l_0 \le x_0 \le L_0-l_0-t-1$, assuming that the dependence on $x_0$ is negligible.
Furthermore, to reduce the amount of data, we binned 1000 measurements, writing
the corresponding averages to the data file.  This allows for a more flexible
analysis of the data.
In particular, we computed the effective correlation length
\begin{equation}
 \xi_{eff}(t) = - \frac{1}{\ln(G_S(t+1)/G_S(t))}
\end{equation}
and correspondingly $c_{eff}(t) = G_S(t) \exp(t/\xi_{eff}(t))$. With
increasing $t$ the effective correlation length approaches $\xi_{exp}$,
which governs the asymptotic behaviour.

In order to reduce errors due to the truncation of the sum at $t_{max}$
in our estimates of the susceptibility and $\mu_2$ we added
in eq.~(\ref{chidefopen}) the contribution
\begin{equation}
\label{extrachi}
 R_{\chi} = 2 \sum_{t=t_{max}+1}^{\infty} c_{eff}(t_{max})  
\exp(-t/\xi_{eff}(t_{max})) \;\;
\end{equation}
and for eq.~(\ref{mu2defopen})
\begin{equation}
\label{extramu2}
 R_{\mu_2} = 2 \sum_{t=t_{max}+1}^{\infty} t^2 c_{eff}(t_{max})  \exp(-t/\xi_{eff}(t_{max})) \;\;,
\end{equation}
which in fact improved the convergence of our estimates of $\chi$ and
$\xi_{2nd}$ with increasing $t_{max}$.  For our final values of
$\chi$ and $\mu_2$  we chose $t_{max}$ such that $\xi_{eff}(t_{max})$ deviates
from $\xi_{exp}$ by a few permille.

We made attempts to fit our data for $G_S(t)$ using a two-exponential Ansatz
\begin{equation}
\label{twoexponential}
 G_S(t) = c_1 \exp(-t /\xi_1) + c_2 \exp(-t /\xi_2) \;\;.
\end{equation}
The results of such fits were not very convincing and we therefore do not
report details. Very roughly our results are consistent with
$\xi_2 \approx \xi_{1}/2$.   For $\xi_2 = \xi_{1}/2$  we tried to get an estimate
for the ratio of amplitudes $c_1/c_2$. We find $c_1/c_2 \approx 9$, $5$,  and $3$
for $N=10$, $21$, and $41$, respectively. Using these numbers, we computed $\xi_{eff}(t)$
for the Ansatz~(\ref{twoexponential}). This allows us to get an estimate for the
deviation of $\xi_{eff}(t)$ from the asymptotic value $\xi_{exp}$.
In order to balance with the statistical error, 
we are aiming at an error of about 2
permille. This is achieved for $t \approx 4 \xi_{exp}$, $4.5 \xi_{exp}$,
and $5 \xi_{exp}$  for $N=10$, $21$, and $41$, respectively.

In the case of the topological charge it turns out that the correlation function
becomes negative for distances larger than zero. The modulus of the correlation
function is decaying fast. This behaviour does not scale with the correlation
length. For example for $N=21$ we find that the effective correlation length of
the topological charge at distance $t=6$ increases from $1.43(14)$ at $\beta=0.675$
up to $2.26(4)$ at $\beta=0.95$. On the other hand, the exponential correlation
length is $3.364(5)$ at  $\beta=0.675$ and $19.147(33)$ at $\beta=0.95$.
Furthermore the effective correlation length of the topological charge clearly
increases with the distance. In our simulations we can not find a plateau.
Therefore an extrapolation similar to eq.~(\ref{extrachi}) is not useful
in the case of the topological susceptibility. Instead we just truncate the
summation at $t_{max}$.

\subsubsection{N=10}
\label{Sopen10}
In table \ref{Oopen10} we summarize the parameters and a few basic results obtained
from our simulations with open boundary conditions for $N=10$. The values of $\beta$
and $L_1$ are chosen such that they match with a subset of
those given in tables 6.2 and 6.3 of ref. \cite{Flynnetal15}. Throughout we started 
with ordered configurations. For $\beta=0.8$ up to $1.0$ we performed $10^6$ update
cycles and measurements. For $\beta=1.02$ and $1.05$ we performed $2 \times 10^6$
and $1.46 \times 10^6$ update cycles, respectively. We discarded up to 
$50000$ measurements for equilibration.
The results for $\chi$ and $\xi_{2nd}$ are computed by using the
extrapolations~(\ref{extrachi},\ref{extramu2})  with 
$t_{max} \approx 4 \xi_{exp}$. The statistical errors of these quantities are
computed by using the Jackknife method. Summing up to $t_{max}=l_0$  without
extrapolation gives consistent results for the magnetic susceptibility.
The error bars are slightly larger in this case.  The results presented for 
the topological susceptibility are obtained by summing the correlation function
of the topological charge up to $t_{max}=l_0$.  
The statistical error depends on the choice of 
$t_{max}$. For example for $t_{max}=l_0/2$, which still might be considered 
as safe,  the statistical error is smaller by  a factor of  $1.5$ up to $1.8$. 

Our results for $E$, $\chi$ and $\chi_t$ are consistent with those of
ref. \cite{Flynnetal15}. In the case of $\xi_{2nd}$ we find a small deviation 
that we attribute to the fact that
different definitions are used. In the case of ref. \cite{Flynnetal15} the 
definition~(\ref{xi2murks}) is used, while we use eq.~(\ref{xi2nd}) along with
eqs.~(\ref{chidefopen},\ref{mu2defopen}). 

\begin{table}
\caption{\sl \label{Oopen10}
Parameters and observables for our runs with open boundary conditions and  $N=10$.
Similar to the tables for the runs with periodic boundary conditions. Here we give 
in addition in column 2 the value of $l_{0}$. 
}
\begin{center}
\begin{tabular}{ccccccccc}
\hline
$\beta$&$L_1$&$l_{0}$&$n_{ov}$& $E$ & $\chi$ & $\xi_{2nd}$ & $\chi_t$ & 
$\tau_{\chi_t}$\\
\hline
0.8 &72 &  44 & 20 & 0.6670240(31)  &   28.0531(29) & 4.6098(17) & 0.0009739(14) & 0.897(6)\\
0.85&96 &  60 & 28 & 0.6222721(21)  &   46.8619(48) & 6.4147(24) & 0.0004617(10) & 1.866(20) \\
0.9 &136&  82 & 38 & 0.5838347(13)  &   78.1989(80) & 8.8441(34) & 0.0002333(8)  & 4.56(9)\\
0.95&184& 112 & 52 & 0.55026611(91) &  130.701(15)  &12.1309(47) & 0.00011930(65)&9.88(23) \\
1.0 & 248&152& 72  & 0.52058941(63) &  219.083(27)  & 16.5884(68)  & 0.00006285(59) &22.0(1.4) \\  
1.02 & 288&174&82  & 0.50964615(37) &  269.527(23)  & 18.7875(54)& 0.00004939(38) &27.4(1.2) \\
1.05 & 344&210&98  & 0.49410845(36) &  368.394(39) & 22.6516(81)& 0.00003414(40) & 36.0(3.7) \\
\hline
\end{tabular}
\end{center}
\end{table}

Since in the case of open boundary conditions a different estimator of the 
topological susceptibility is used as for periodic boundary conditions, it  
is not sufficient to compare the integrated autocorrelation times 
of the simulations. Instead we define a performance index as 
the inverse of the square of the relative statistical 
error of the topological susceptibility divided by the number of the sweeps. 
In our case, the number of the sweeps is given
by stat $\times (1+n_{ov})$. Furthermore, we multiply by 4, to take the larger area of 
the lattice into account.   Results for the performance index are summarized
in table \ref{comp10}. At small values of $\beta$, where the 
integrated autocorrelation time of the topological susceptibility is still 
small for periodic boundary conditions, the simulation with open 
boundary conditions
is less efficient than that with periodic ones. At about $\beta=0.9$
we find roughly the same performance, while going to even lager values of 
$\beta$, the open boundary conditions become more and more favourable. 
This comparison depends on the choice of $t_{max}$ in the case of open
boundary conditions. Taking $t_{max}=l_0/2$ instead of $t_{max}=l_0$ 
we get an additional factor of $1.5^2$ up to $1.8^2$ in advantage of 
open boundary conditions.

Since for $N=10$ standard simulations with periodic boundary conditions are 
still feasible for a rather large correlation length, we could not demonstrate 
a really decisive advantage for open boundary conditions. The situation is 
different for the larger values of $N$ discussed below. 

\begin{table}
\caption{\sl \label{comp10}
We give the performance index as defined in the text for simulations
with periodic boundary conditions based on the numbers given in table 6.3
of ref. \cite{Flynnetal15}  and for open boundary conditions based on 
the results of our simulations.  $N=10$.
}
\begin{center}
\begin{tabular}{cllc}
\hline
$\beta$ &  periodic  & open &  ratio \\
\hline
  0.8   &   0.00973  &  0.00576 & 0.6 \\
  0.85  &   0.00221  &  0.00184 & 0.8 \\
  0.9   &   0.000555 &  0.000545& 1.0 \\
  0.95  &   0.000118 &  0.000159&  1.3 \\
  1.0   &   0.0000217&  0.0000389 &  1.8 \\ 
  1.02  &   0.00000864&  0.0000254 & 2.9 \\
  1.05  &   0.00000385&  0.0000126 & 3.3 \\
\hline
\end{tabular}
\end{center}
\end{table}

\subsubsection{N=21}

In table \ref{Parameter21} we summarize the parameters and a few basic results obtained
from our simulations with open boundary conditions for $N=21$. The analysis 
is performed in a similar fashion as for $N=10$. 
Our results 
for $E$, $\chi$ and $\chi_t$ are consistent with those given in tables A and B
of ref. \cite{Metadynamics}.  The small difference in $\xi_{2nd}$ can be 
attributed to the different definitions that have been used.  Note that
for $\beta=0.85$  the authors of ref. \cite{Metadynamics} find 
$\chi_t = 0.000042(23)$. Our result $\chi_t = 0.00004421(77)$ is fully 
consistent. However our error bar is smaller by a factor of 30. 

\begin{figure}
\begin{center}
\includegraphics[width=15.0cm]{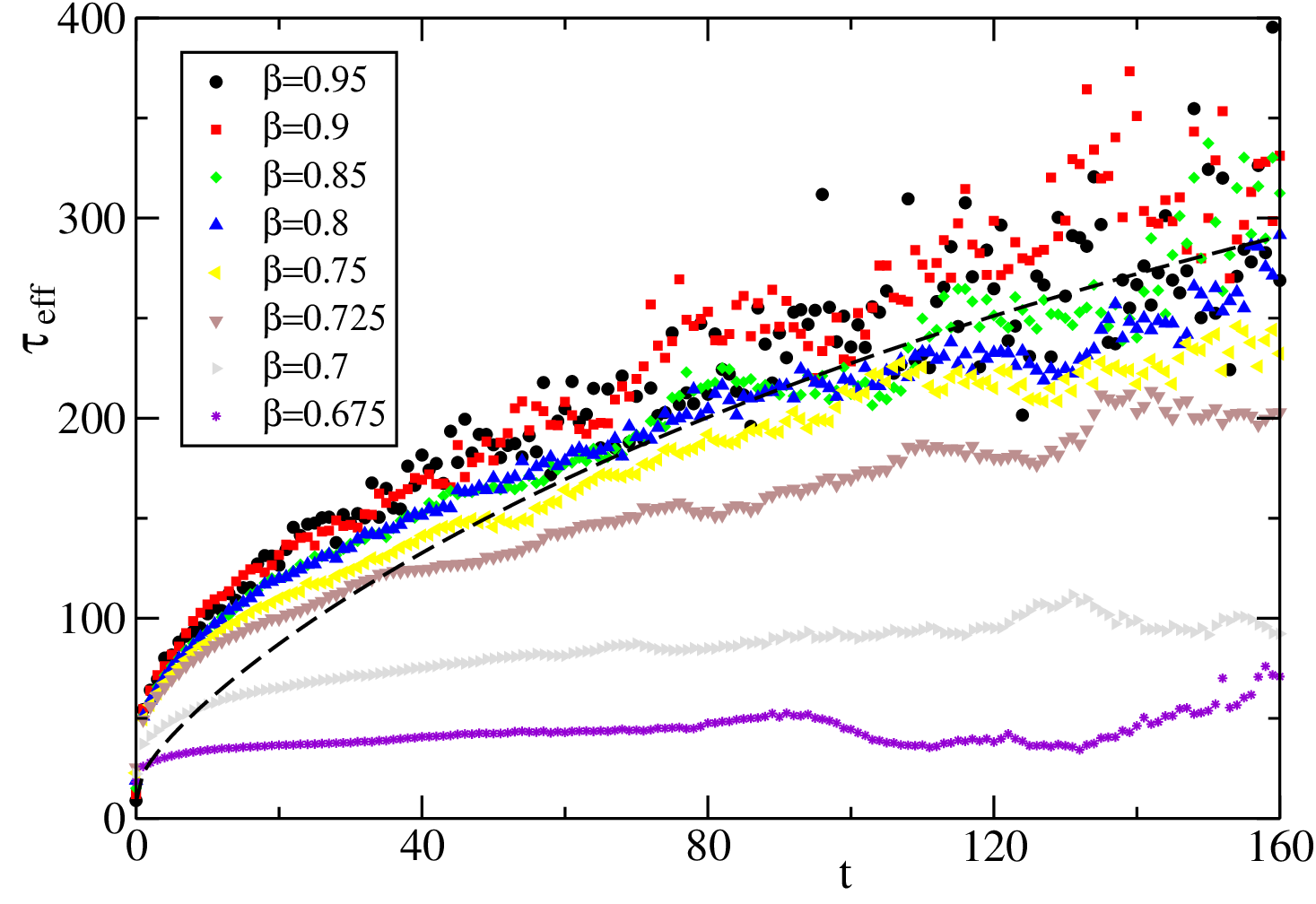}
\caption{\label{tauopen21}  We plot the effective autocorrelation time of
the topological susceptibility
$\tau_{eff,10,\chi_t}$ as a function of the distance $t$ in the Markov chain.
The data are taken from the simulations with open boundary
conditions and $N=21$. To keep the plot readable we skipped the error bars.
The dashed line gives the effective autocorrelation time obtained from 
the Ansatz~(\ref{hypothetical}) with $\tau_{exp} =455$. 
}
\end{center}
\end{figure}

\begin{table}
\caption{\sl \label{Parameter21}
Parameters and observables  of the runs for $N=21$ and open boundary conditions.
Similar to table \ref{Oopen10} for $N=10$.
Throughout $2 \times 10^6$ update cycles are performed. We discarded
up to $50000$ measurements at the beginning of the simulation.
}
\begin{center}
\begin{tabular}{ccccccccc}
\hline
$\beta$&$L_1$&$l_{0}$&$n_{ov}$& $E$ & $\chi$ & $\xi_{2nd}$ & $\chi_t$ & $\tau_{int,\chi_t}$ \\ 
\hline
0.625 &  38 &  22 & 3  & 0.8333946(36)  &  9.6087(5)  &  2.2968(5) & 0.0017056(50) & 8.32(11) \\ 
0.65  &  44 &  26 & 4  & 0.7993498(29)  & 12.1470(7)  &  2.7137(5) & 0.0011668(46) & 14.95(28)\\
0.675 &  52 &  32 & 5  & 0.7680222(24)  & 15.3903(9)  &  3.1992(7) & 0.0008015(47) & 29.4(1.1) \\
0.7   &  60 &  38 & 6  & 0.7390678(20)  & 19.5466(12) &  3.7638(8) & 0.0005708(48) & 57.2(2.9) \\
0.725 &  72 &  44 & 7  & 0.7122223(15)  & 24.8775(16) &  4.4188(10)& 0.0004220(50) & 117.(8.) \\
0.75  &  84 &  52  & 8 & 0.6872687(13)  & 31.7435(20) &  5.1852(12)& 0.0002891(40) & 146.(12.) \\
0.8   & 112 &  70 & 11 & 0.6422831(9)   & 51.9924(34) &  7.1207(16)& 0.0001534(23) & 154.(13.) \\
0.85  & 156 & 100 & 15 & 0.6028740(6)   & 85.7964(58) &  9.7556(23)& 0.0000803(13) & 145.(13.) \\
0.9   & 212 & 132 & 20 & 0.5680773(4)   & 142.539(10) & 13.3544(33)& 0.00004421(77)& 156.(15.) \\
0.95  & 300 & 182 & 28 & 0.5371311(3)   & 238.187(16) & 18.2419(43)& 0.00002327(37)& 103.(9.)  \\
\hline
\end{tabular}
\end{center}
\end{table}
In Fig. \ref{tauopen21} we give the effective autocorrelation time
of the topological susceptibility $\tau_{eff,10,\chi_t}$  
for $N=21$ and open boundary conditions. 
We computed $\tau_{eff,s,\chi_t}$ also for different
values of $s$. We find that the behaviour of $\tau_{eff,s,\chi_t}$  depends
little on $s$. For larger $s$ the values of $\tau_{eff}$ scatter less.

For $\beta=0.675$ the value of $\tau_{eff,10,\chi_t}$ still levels off 
reasonably well with increasing $t$.  We read off $\tau_{exp} \approx 50$, 
which is similar to $\tau_{int}=63.1(2.5)$ that we get from the simulation
with periodic boundary conditions. For $\beta=0.7$ one would read off 
$\tau_{exp} \approx 100$, which is well below $\tau_{int}=199.(18.)$ that
we find for periodic boundary conditions. For $\beta=0.725$ one can 
hardly spot a plateau.  But it is quite plausible that we stay well below 
$\tau_{int,\chi_t}=524.(64.)$ obtained for periodic boundary conditions.  
Starting
from $\beta=0.75$ we can not find a plateau in the range that is plotted.

The other interesting observation is that for increasing $\beta$ 
the curves for $\tau_{eff,10,\chi_t}$ seem to fall on top of each other
for different values of $\beta$. This means that slowing down is eliminated 
up to the increasing number of overrelaxation steps $n_{ov} \propto \xi$
per measurement.  Correspondingly we find that the integrated autocorrelation
times given in table  \ref{tauopen21} stay roughly constant starting 
from $\beta=0.75$. For $\beta=0.95$ we find an even smaller value of 
$\tau_{int,\chi_t}$ again. This is mainly due to the behaviour at large 
distances $t$ in the Markov chain. It is not clear to us, whether this is 
of significance.

Following the  model for the diffusion of instantons in the lattice
presented in ref. \cite{McGlMa14}, the autocorrelation function takes the form 
\begin{equation}
\label{hypothetical}
 \rho(t) \propto \sum_n   c_n  \exp\left(-\frac{n^2}{\tau_{exp}} t  \right) \;,
\end{equation}
when the dynamics of the instantons is dominated by diffusion.  We made no 
attempt to work out the coefficient $c_n$ based on the diffusion model. 
In Fig. \ref{tauopen21} we plot as dashed line the effective autocorrelation 
time computed from eq.~(\ref{hypothetical}), where we took $c_n = n^{-1}$ 
and $\tau_{exp}=455$. There is a reasonable match with $\tau_{eff}$ obtained 
from the data for $\beta=0.85$ for $t \gtrapprox 80$.

\begin{figure}
\begin{center}
\includegraphics[width=15.0cm]{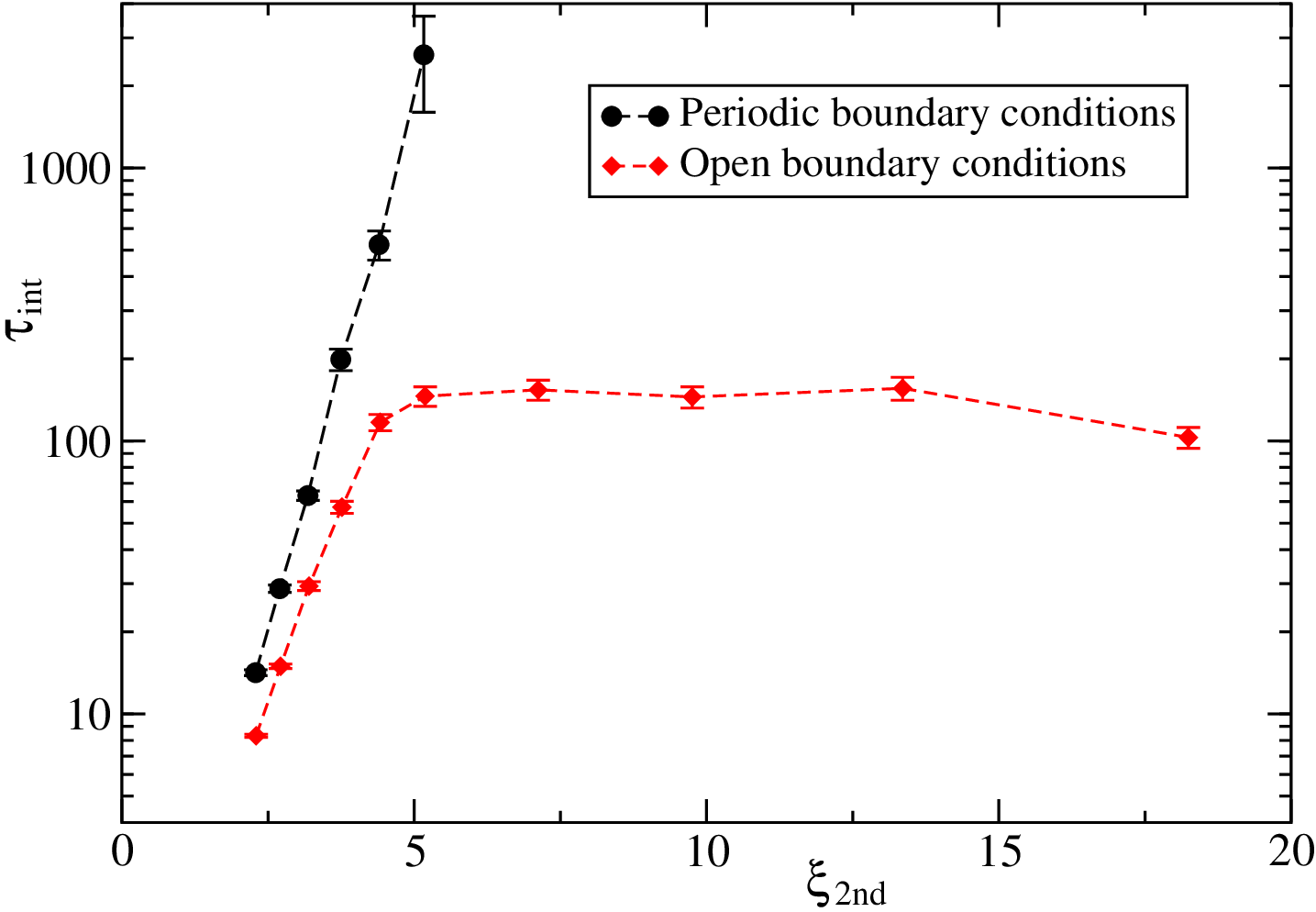}
\caption{\label{tauV21}  We plot the integrated autocorrelation time of
the topological susceptibility
$\tau_{int,\chi_t}$ for $N=21$ as a function of the second
moment correlation length $\xi_{2nd}$. We give results for our  
standard simulations with periodic boundary conditions in both directions
and for open boundary conditions in $0$-direction. For a discussion 
see the text.
}
\end{center}
\end{figure}

Finally, in Fig. \ref{tauV21} we compare the integrated autocorrelation of the 
topological susceptibility for open and periodic 
boundary conditions in $0$-direction. The numbers are taken from table 
\ref{P21} and \ref{Parameter21}, respectively. As already discussed in section 
\ref{standardperiodic}, the autocorrelation time seems to increase 
exponentially with the correlation length in the case of periodic boundary
conditions.
Instead, for open boundary conditions, we first see an increase that is 
similar to that with periodic boundary conditions. 
Then, at $\xi_{2nd} \approx 5$ the autocorrelation time levels off.
The difference between open and periodic boundary conditions for smaller
values of $\xi_{2nd}$ is mainly due to the fact that different estimators
of the topological susceptibility are used.  The behaviour in the case 
of open boundary conditions can be explained along the lines of ref.
\cite{McGlMa14}. For $\xi_{2nd} \lessapprox 5$ changes of the topological
charge are dominantly due to the creation and destruction of instantons 
in the bulk. Then for $\xi_{2nd} \gtrapprox 5$ the diffusion from and to 
the boundaries completely dominates. This diffusion is not effected by 
the severe slowing down.

\subsubsection{N=41}
In table \ref{Parameter41} we summarize the parameters and a few basic 
results obtained from our simulations with open boundary conditions
for $N=41$. Note that all four values of $\beta$ are in the range, where 
standard simulations with periodic boundary conditions show topological 
freezing. We abstain from a detailed discussion of the simulations, since
the findings are very similar to those for $N=21$.  

\begin{table}
\caption{\sl \label{Parameter41}
Parameters and observables of the runs for $N=41$ and open boundary conditions. 
Similar to table \ref{Oopen10} for $N=10$. 
Throughout $2 \times 10^6$ update cycles are performed. We discarded 
$50000$ measurements at the beginning of the simulation. 
}
\begin{center}
\begin{tabular}{ccccccccc}
\hline
$\beta$ & $L$ &$l_0$& $n_{ov}$  & $E$  & $\chi$  & $\xi_{2nd}$ & $\chi_t$ & $\tau_{int,\chi_t}$ \\
\hline
0.65 &  56 & 34 & 6 & 0.7810290(15) & 15.9354(6)   & 3.3730(6)&0.0003598(51) & 173.(13.)  \\
0.7  &  76 & 48 & 8  & 0.7249816(10)   &  25.5383(11)  & 4.6324(8) &0.0001858(28) & 174.(15.)  \\
0.75 & 104 & 64& 11  & 0.6760887(7) & 41.4379(17)  & 6.3550(11) &0.0000977(15) & 171.(16.) \\
0.8  & 140 & 86& 15  & 0.6331986(5) & 67.9135(30)  & 8.7023(16) &0.0000518(8)  & 145.(13.) \\
\hline
\end{tabular}
\end{center}
\end{table}

Note that the integrated autocorrelation times of the topological susceptibility are only slightly
larger than those for $N=21$ and $\beta \ge  0.75$. 

\subsection{Periodic boundary conditions: parallel tempering}
\label{NumTempering}
\subsubsection{N=10}
We started our numerical experiments with open boundary conditions. We
performed simulations at $\beta=1.05$, where 
the autocorrelation time of the topological susceptibility is already 
large.  We have chosen $L=L_0=L_1=344$ as linear size of the lattice. 
We performed a number of preliminary
simulations to get an idea how the parameters of the update scheme
should be chosen. Let us report the details of our final choice. 
The number of replica is $N_t=96$. We performed 6000 update cycles.
For swaps between $t=0$ and $1$ this 
gives an acceptance rate of about $80 \%$ and drops to a bit more than 
$35 \%$ for the pair $t=75$ and $76$. Then it increases again up to 
a little less than $52 \%$ for $t=94$ and $95$. We have taken $l_1=32$
and $n_1=6$, $n_2=n_3=n_4=n_5=n_6=2$ as parameters of the cycle.
As in the simulations with open boundary conditions, we used $n_{ov}=98$
for the sweeps over the whole lattice. For levels  $1, 2, 3, 4, 5,$ and $6$
 we used 
$n_{ov}=98, 49, 24, 24, 24$, and $24$, respectively.  Here we  performed 
measurements along with the updates at level 4 of the update cycle.

The efficiency of the parallel tempering algorithm can be characterized
by the round trip time. This means that one follows the configurations 
on their way through the parameter of the tempering scheme. Based on this,
one computes the average time that is needed to go from one end of the 
chain to the other.  In our simulations, we followed only a single 
configuration to reduce the amount of data that is generated. For the 
same reason, the position of the configuration is recorded only once per 
cycle. This way, we might miss arrivals of the configuration at the 
top or the bottom of the tempering chain.  Therefore we just computed the 
integrated autocorrelation time of the position.  As average position 
in the chain we get $t_{av}=49.1(1.0)$, which is in reasonable agreement
with the exact result $95/2=47.5$. The integrated autocorrelation is
$\tau_{int,Pos}=3.6(5)$, indicating that the configurations are indeed 
shuffled through the tempering chain within a few update cycles. Note that
within one cycle 379 swap updates of the configurations are performed.

In the analysis of the data we discarded the first 1000 update cycles.
We get $E=0.4941087(16)$ and $\chi=368.31(20)$ for the energy and the 
susceptibility, respectively. 
For these two observables the autocorrelation almost 
vanishes. The integrated autocorrelation times are $\tau_{int}=0.57(2)$ 
and $0.59(2)$, respectively. For the topological susceptibility we
get $\chi_t=0.00003344(42)$ and  $\tau_{int,\chi_t}=3.0(4)$.  These numbers can 
be compared with $E=0.49410872(8)$, $\chi=368.33(5)$ and $\chi_t=0.0000340(9)$
given in table 6.3 of ref. \cite{Flynnetal15}  and 
$E=0.49410845(36)$, $\chi=368.394(39)$ and 
$\chi_t=0.00003414(40)$ obtained in our simulation with open boundary
conditions presented in section \ref{Sopen10} above. Taking into account
the reduced number of $n_{ov}$ at higher levels of the update scheme, 
the numerical cost of this run is by a factor of $2.6$ less than 
that of our simulation for $\beta=1.05$ with open boundary conditions.
This means that, looking only at the topological susceptibility, there 
is a small advantage for the parallel tempering run. 

Next we studied $l_d < L_1$. The disadvantage  of this choice is that 
less topological objects can be created or destroyed in a unit of time.
On the other hand smaller values of $N_t$ should be sufficient to 
give reasonable acceptance rates for the swaps of configurations between
different $t$.  This is in particular important with regard to the 
application to lattice QCD.  In addition to saving memory space, smaller values
of $N_t$ should also allow for smaller round trip times.  After a few 
preliminary simulations, where we mainly determined acceptance rates for 
the swaps of configurations, we performed extended simulations for 
$\beta=1.05$, $L=L_0=L_1=344$ and defect lines characterized by $l_d=8, 16$,
and $24$.  The number of replica is $N_t=24$ in all three cases. 
The update scheme is characterized by $l_1=32$ in all three cases
and $n_1=23$, $n_2=4$, $n_3=n_4=n_5=3$ and $n_6=2$ for $l_d=8$,
$n_1=23$, $n_2=n_3=n_4=3$ and $n_5=n_6=2$ for $l_d=16$, and
$n_1=21$, $n_2=n_3=n_4=3$ and $n_5=n_6=2$ for $l_d=24$.
In total we performed 25000, 60000, and 42000 update cycles for $l_d=8$,
$16$, and $24$, respectively.
For $l_d=8$ the acceptance rate is about $87.3 \%$ for the pair $t=0$ and $1$. 
It drops to  $58.0 \%$ for $t=20$ and $21$ and goes slightly
up again to $58.9 \%$ for $t=22$ and $23$.  For $l_d=16$ 
the acceptance rate is about $82.0 \%$ for the pair $t=0$ and $1$. It drops
to  $44.2 \%$ for $t=18$ and $19$ and increases again to $49.7 \%$ 
for $t=23$ and $24$.  For $l_d=24$ 
the acceptance rate is about $79.9 \%$ for the pair $t=0$ and $1$.
It drops to  $36.1 \%$ for $t=19$ and $20$ and increases again to $44.0 \%$
for $t=23$ and $24$.  
As expected, for fixed $N_t$ the acceptance rates for the swaps of 
configurations decrease with increasing length of the defect line.  
Preliminary runs and the run with open boundary conditions suggest that
\begin{equation}
 N_t \propto \sqrt{l_d}
\end{equation}
gives an acceptance rate that is roughly constant in $l_d$. In these 
simulations we have written the value of $t$ for one given configuration
to the file when performing the update at level 2. This way translational 
invariance in Monte Carlo time is lost. In the following analysis we shall
ignore this fact when computing autocorrelation times of the position. 
In units of complete cycles, we get $\tau_{int,Pos}=0.574(2)/92, 
1.102(5)/69$, and $1.730(13)/63$ for $l_d=8$, $16$, and $24$, respectively.
This means that within a single cycle configurations move several times
from the top to the bottom of the tempering chain. For the topological 
susceptibility we get $\chi_t=0.00003386(32)$, $0.00003389(16)$, and
$0.00003385(16)$, respectively. The integrated autocorrelation times
of the topological susceptibility are $\tau_{int,\chi_t}=5.9(6)$,  $3.41(20)$, 
and $2.46(15)$, respectively. We see no sharp dependence of the performance
on $l_d$. The best performance is achieved for $l_d=24$, suggesting 
that $l_d \approx \xi$ is a good choice. Our run with $l_d=L_1$ is clearly 
outperformed by $l_d=24$. 

\subsubsection{$N=21$}
Here we performed simulations at $\beta=0.8$, $0.85$, $0.9$ and $0.95$, 
where standard simulations fail to equilibrate.  The parameters of our
simulations and the numerical results for the energy density, the magnetic 
susceptibility and the topological susceptibility 
are summarized in table \ref{Parameter21Temper}. We performed 200000, 200000,
141800, and 50370 update cycles, respectively.
Let us discuss the simulation
at $\beta=0.95$ in more detail. The update cycle is characterized by
$n_1=24$, $n_2=n_3=n_4=3$, and $n_5=n_6=2$. The number of overrelaxation 
updates per cycle is $28$, $14$, $7$, $7$, $3$, and $3$ at levels $1$, $2$, $3$, $4$,
$5$, and $6$, respectively. The acceptance rates for the swaps
of configurations behave similar to the $N=10$ case.
We find that the acceptance rate is about $81.4 \%$ 
for the pair $t=0$ and $1$. It drops to  $39.4 \%$ for the pair 
$t=26$ and $27$. Then it increases again to $47.3 \%$ for the pair $t=30$ and 
$31$.  The integrated autocorrelation time of the position
of a selected configuration is $\tau_{int,Pos}= 1.875(13)/72$. 
Remember that we always write to the file at level 2 of the update cycle,
which means 72 times per cycle in the present case. The simulation took 25 days on 
a 4 core PC running with 8 threads.  This is about the same CPU time 
that is used for the corresponding run with open boundary conditions. The 
error bar of the topological susceptibility is smaller by a factor of $2.3$
compared with the simulation with open boundary conditions.

\begin{table}
\caption{\sl \label{Parameter21Temper}
Parameters of the runs for $N=21$ using parallel tempering.  In addition
we give results for the energy density, the magnetic susceptibility, 
the topological susceptibility and the integrated autocorrelation time
of the topological susceptibility.
}
\begin{center}
\begin{tabular}{cccccccccc}
\hline
$\beta$ & $L$ & $n_{ov}$ & $l_d$ & $l_1$ & $N_t$ & $E$ & $\chi$ & $\chi_t$ & 
$\tau_{int,\chi_t}$\\
\hline
0.8 &112 &11 & 8 & 16 & 16 & 0.6422829(12)& 51.9872(46)&0.00015574(58)&4.28(15) \\
0.85&156&15  & 8 & 16 & 16& 0.6028732(7) & 85.793(7)  & 0.00008196(39) & 9.6(5) \\
0.9& 212& 20 &12& 32 & 24& 0.5680766(5) & 142.506(13) & 0.00004377(18) & 5.5(2)  \\
0.95&300 &28 &16 &32 & 32& 0.5371312(6)& 238.223(36) &0.00002305(16) & 7.2(5) \\
\hline
\end{tabular}
\end{center}
\end{table}

\subsubsection{$N=41$}
Finally we performed a simulation for $N=41$ at $\beta=0.8$, which is the 
largest value of $\beta$ that we simulated with open boundary conditions.
We simulated a $140^2$ lattice with $l_d=16$ and $N_t=32$.  
Performing 200000 update cycles, we get $E=0.633198(6)$, $\chi=67.9260(29)$,
$\chi_t=0.00005169(21)$ and $\tau_{int,\chi_t} = 6.3(3)$. 
Also here we find that the parallel tempering simulation is more efficient
than the simulation with open boundary conditions  by a factor of about $5$
when focussing on the topological susceptibility.

\section{Physics results and comparison with the large $N$-expansion}
Finally we try to extract results for the continuum limit from our data 
and compare these  with predictions 
obtained from the $1/N$-expansion \cite{ML78,CaRo91,CaRo93}.
For the correlation length and the susceptibility we shall  use the data 
obtained from our simulations with open boundary conditions. In the 
case of the topological susceptibility we use the results obtained from
our simulations with parallel tempering when available and the 
results obtained from our simulations with open boundary conditions otherwise. 

\begin{figure}
\begin{center}
\includegraphics[width=14.5cm]{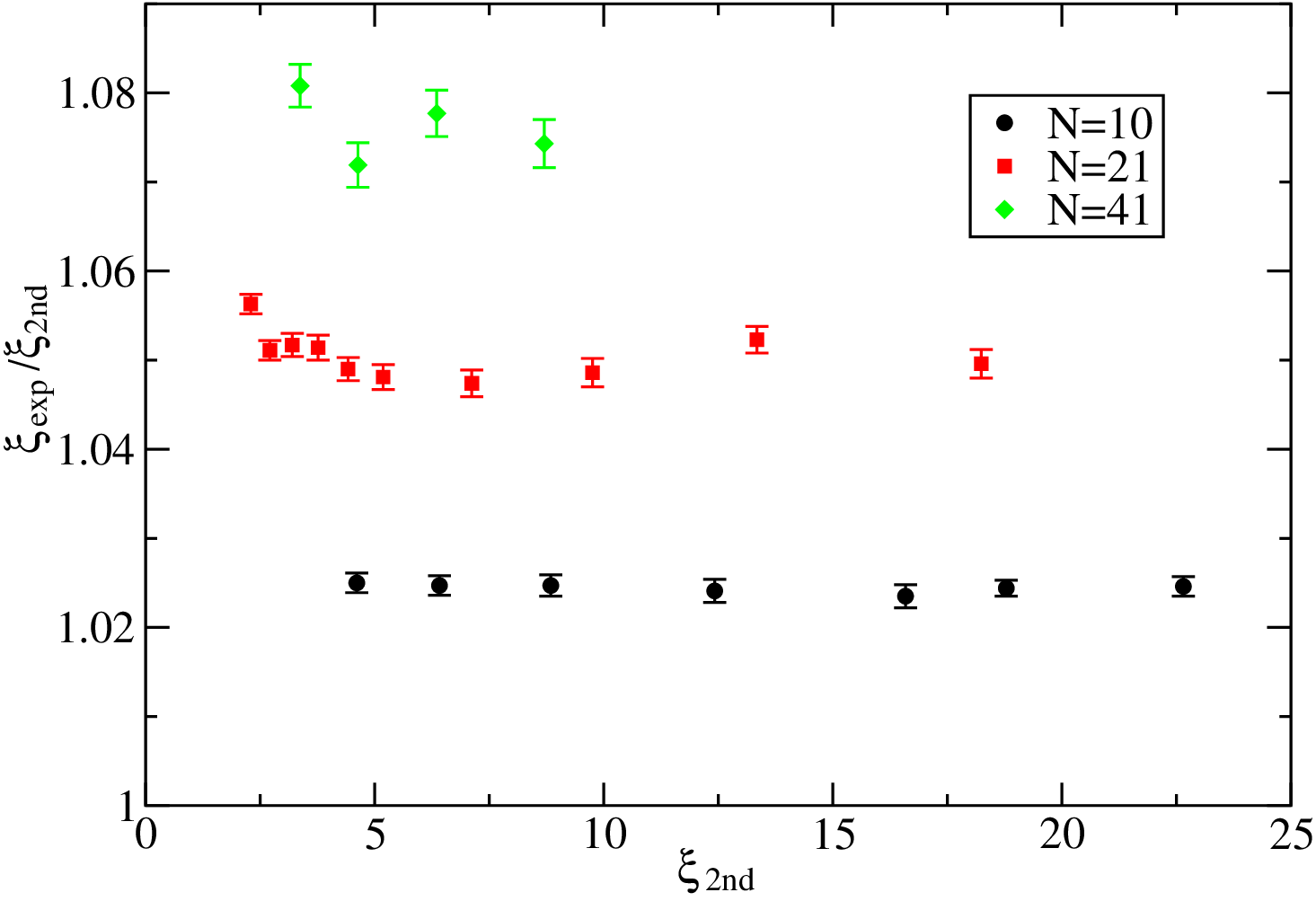}
\caption{\label{plotRxi}  We plot the ratio $\xi_{exp}/\xi_{2nd}$ for
$N=10$, $21$ and $41$ as a function of $\xi_{2nd}$.}
\end{center}
\end{figure}
Let us first consider the ratio of second moment and the exponential 
correlation length. Our numerical
results are presented in Fig. \ref{plotRxi}. For all values of $N$
that we studied, no clear dependence on $\xi_{2nd}$ can be seen.
Our final results are $\xi_{exp}/\xi_{2nd}=1.024(3), 1.049(3)$,
and $1.076(4)$ for $N=10$, $21$ and $41$,
respectively. These were obtained by averaging  all
results for $\xi_{2nd} > 5$. The quoted error takes into account the statistical
error as well as the systematical two permille error of $\xi_{exp}$. Our 
results are consistent with those given in tables X and XV of 
ref. \cite{campostrini}.
Following ref. \cite{CaRo91}
\begin{equation}
 \frac{\xi_{2nd}}{\xi_{exp}} = \sqrt{\frac{2}{3}} + O\left(N^{-2/3}\right) \;.
\end{equation}
Our results obtained for $N=10$, $21$ and $41$ are still quite far from
this asymptotic value. Therefore we abstain from estimating the coefficient of
the  $O(N^{-2/3})$ corrections.

\begin{figure}
\begin{center}
\includegraphics[width=14.5cm]{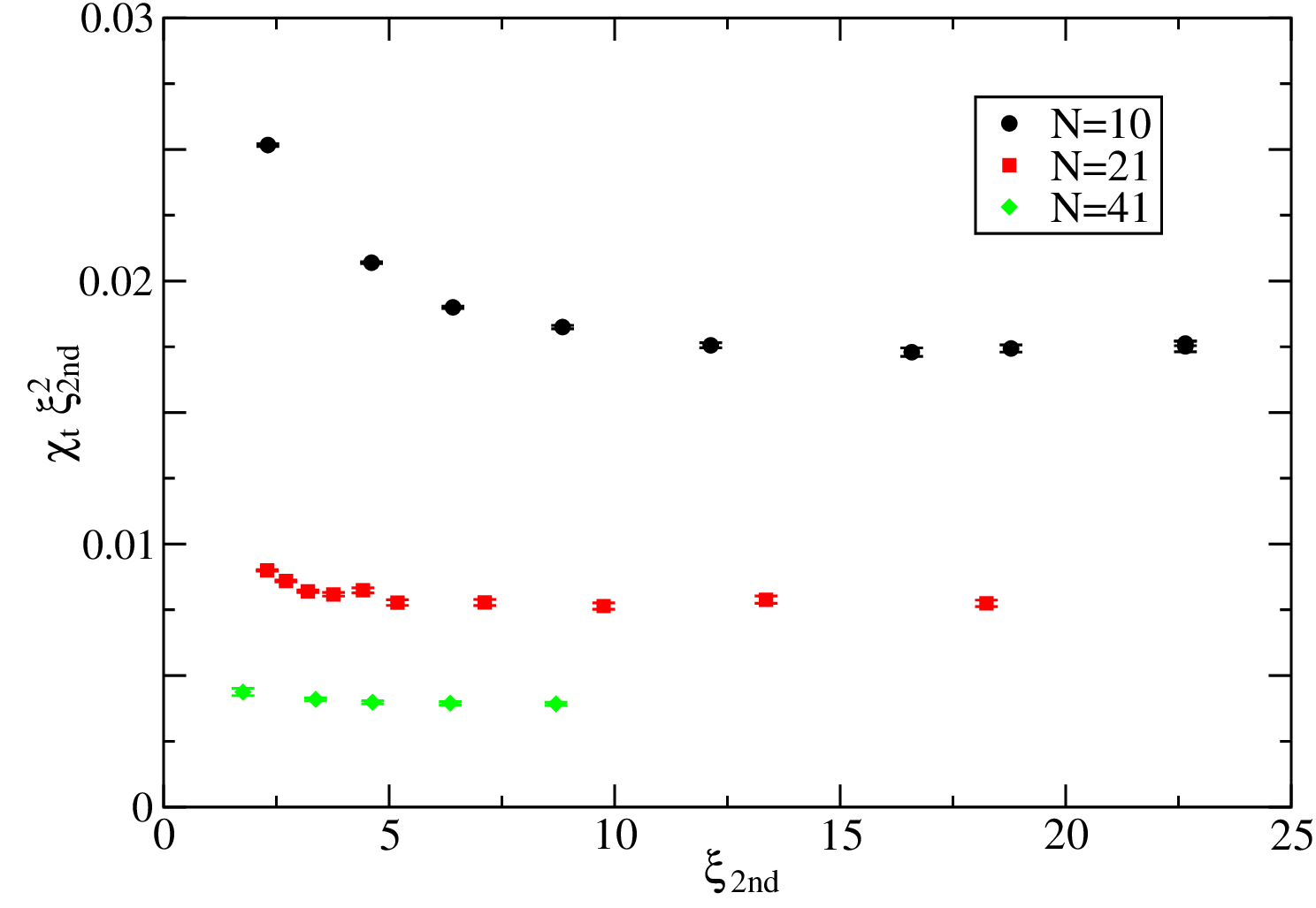}
\caption{\label{plotxi2chit}  We plot $\xi_{2nd}^2 \chi_t$ as a function
of $\xi_{2nd}$ for $N=10$, $21$ and $41$.}
\end{center}
\end{figure}

Next we study the relation between the topological susceptibility and 
the correlation length. The product $\chi_t \xi^2$ should  have a finite 
continuum limit. For the exponential correlation length the $1/N$-expansion 
gives \cite{ML78}
\begin{equation}
\chi_t \xi_{exp}^2 = \frac{3}{4 \pi N} + O\left(N^{-5/3}\right) \;\;. 
\end{equation}
For the second moment correlation length a faster convergence with increasing 
$N$ is obtained \cite{CaRo91}
\begin{equation}
\label{chitxi}
\chi_t \xi_{2nd}^2 = \frac{1}{2 \pi N} \left(1 - \frac{0.38088...}{N}  \right) +  O\left(N^{-3}\right) \;.
\end{equation}
In Fig. \ref{plotxi2chit} we plot $\xi_{2nd}^2 \chi_t$ as a function
of $\xi_{2nd}$.  Looking at the figure, the numerical data seem to converge 
nicely to the scaling limit.  Corrections to scaling seem to be smaller for 
larger values of $N$. 
Taking simply the largest values of $\beta$ for each $N$ we get 
$\xi_{2nd}^2 \chi_t = 0.01737(8)$, $0.00767(5)$ and $0.00391(2)$ for
$N=10$, $21$ and $41$, respectively. 
This can be compared with results quoted in the literature. 
For $N=10$ one finds for example 
$\xi_{2nd}^2 \chi_t = 0.01719(10)(3)$ and $0.0175(3)$ in 
refs. \cite{Flynnetal15,Vicari04}, respectively. For $N=21$ one finds
$\xi_{2nd}^2 \chi_t =0.0080(2)$ and $0.0076(3)$  in refs. 
\cite{Vicari04,Vicari92}, respectively. For $N=41$ we find in 
ref. \cite{Vicari92} the results 
$\xi_{2nd}^2 \chi_t = 0.0044(4)$ and $0.0036(4)$ for
$\beta=0.57$ and $0.6$, respectively. Our estimates are 
essentially consistent with those presented in the literature.  In particular 
for large values of $N$, we improved the accuracy of the estimates. 
To see the effect of leading corrections, it is useful to multiply
$\xi_{2nd}^2 \chi_t$ by $2 \pi N$. Using our numbers, we get $1.091(5)$, 
$1.012(7)$, and $1.007(5)$ for $N=10$, $21$, and $41$, respectively. 
As already discussed in ref. \cite{Vicari04} it is a bit puzzling that 
the numbers suggest a $1/N$ correction with the opposite sign as that of 
eq.~(\ref{chitxi}).
 
\begin{figure}
\begin{center}
\includegraphics[width=14.5cm]{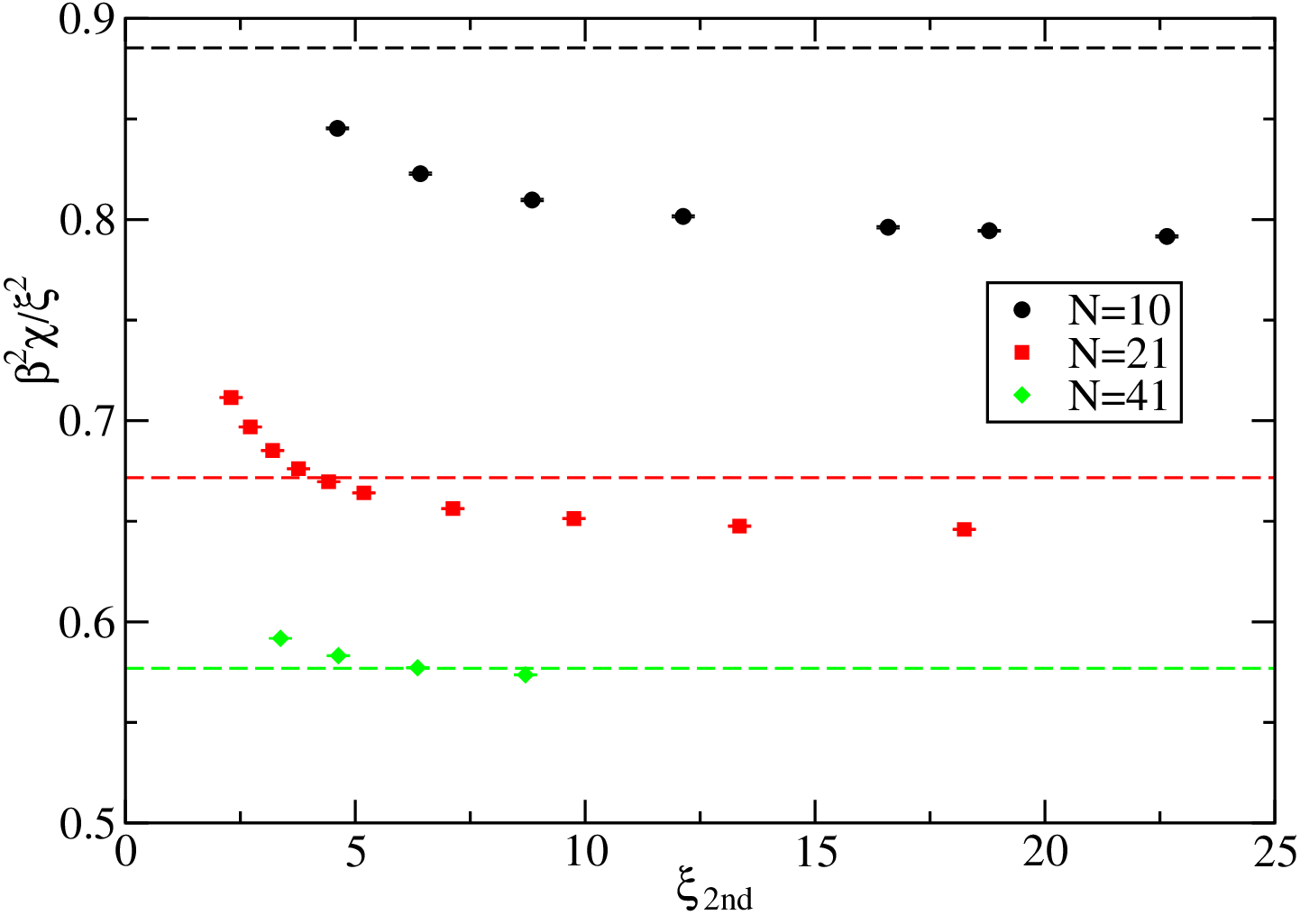}
\caption{\label{plotb2cx2}  We plot the ratio $\frac{\beta^2 \chi}{\xi_{2nd}^2}$,eq.~(\ref{ratiowithb}), as a function of the second moment correlation 
length $\xi_{2nd}$ for $N=10$, $21$ and $41$.
For comparison we give the $O(1/N)$ estimates~(\ref{ratiowithb},\ref{coeff})
obtained from the $1/N$-expansion as dashed lines.}
\end{center}
\end{figure}

In  refs. \cite{campostrini,Vicari92} the ratio $Z_P = \chi \xi_{2nd}^{-2}$, 
which is related to the renormalization of the composite operator $P$, is 
discussed. Following eq.~(8) of the preprint version of ref. \cite{Vicari92}
\begin{equation}
\label{ratiowithb}
\beta^2 Z_p= \frac{\beta^2 \chi}{\xi_{2nd}^2}  = 
c \left[1 + O\left(\frac{1}{N \beta} \right) \right] 
\end{equation}
with 
\begin{equation}
\label{coeff}
c =\frac{3}{2 \pi} \left[1 + \frac{8.5414}{N} + O\left(\frac{1}{N^2} \right) \right]  \;\;.
\end{equation}
In Fig. \ref{plotb2cx2} we plot $\beta^2 Z_p$ as a function of $\xi_{2nd}$.
The estimates seem to converge with increasing $\xi_{2nd}$. However fitting 
the data with Ans\"atze that contain $1/\beta$ and $1/\beta^2$ corrections
show that reliable estimates for the limit $\beta \rightarrow \infty$ can 
hardly be obtained. Note that our data cover only a rather small range of 
$\beta$. Fits with $1/\beta$ only, give large values of $\chi^2/$d.o.f. .
Adding a $1/\beta^2$ correction, the quality of the fit improves and the 
amplitude of the $1/\beta$ correction becomes negative. As a result, the 
estimate of $\lim_{\beta \rightarrow \infty} \beta^2 Z_p$ obtained from these
two Ans\"atze differ strongly. Hence we refrain from giving a final result.
In order to get reliable results, a larger range in $\beta$ has to be 
covered. To this end, finite size scaling approaches  as discussed in refs. 
\cite{running,CaEdFePeSo95} are needed.
In order to get a rough idea how our numerical results compare with those
of the $1/N$ expansion, we give in Fig.  \ref{plotb2cx2} the $O(1/N)$ estimates
of $\lim_{\beta \rightarrow \infty} \beta^2 Z_P$ as dashed lines. Given the
fact that the extrapolation of our data to $\beta \rightarrow \infty$ is 
difficult and $O(1/N^2)$ in the $1/N$ expansion are not known, the numerical 
data and the results of the $1/N$ expansion seem to be consistent.

\begin{figure}
\begin{center}
\includegraphics[width=14.5cm]{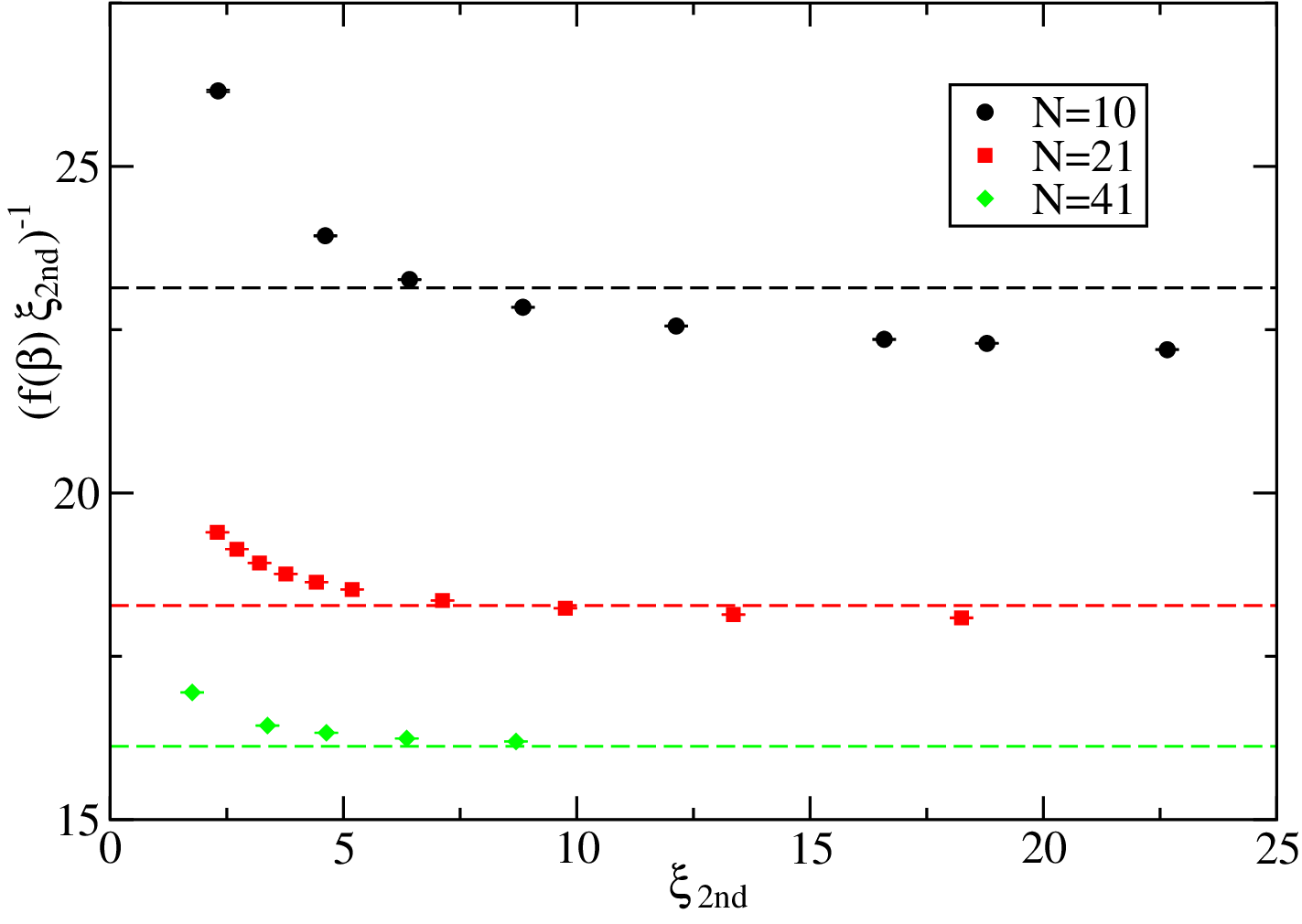}
\caption{\label{plotAscaling} We plot $[\xi_{2nd} f(\beta)]^{-1}$ as 
a function of $\xi_{2nd}$ for $N=10$, $21$ and $41$.
For comparison we give the $O(1/N)$ estimates~(\ref{lambdalargeN})
obtained from the $1/N$-expansion as dashed lines.}
\end{center}
\end{figure}

The asymptotic scaling behaviour of the correlation length is governed by 
the $\beta$-function. In particular 
\begin{equation}
 M_{2nd} /\Lambda_L = \lim_{\beta \rightarrow \infty} [\xi_{2nd} f(\beta)]^{-1}
\; ,
\end{equation}
where the two-loop $\beta$-function implies
\begin{equation}
 f(\beta) = (2 \pi \beta)^{2/N} \; \exp(-2 \pi \beta) \;\;.
\end{equation}
In Fig. \ref{plotAscaling} we plot our numerical results for 
$1/[\xi_{2nd} f(\beta)]$. Also here one expects that corrections vanish 
with powers of $\beta^{-1}$. Therefore it is virtually impossible to 
extrapolate our numerical results to $\beta \rightarrow \infty$. To overcome
this problem, finite size scaling methods as discussed in refs. 
\cite{running,CaEdFePeSo95} are needed.

Despite this problem let us compare 
our estimates with the predictions of the $1/N$-expansion. 
Following ref. \cite{CaRo92}, for the explicit number of the $1/N$ coefficient see eq.~(65) of ref. \cite{VicariRossi93},  
\begin{equation}
\label{lambdalargeN}
\frac{M_{2nd}}{\Lambda_L} = 8 \sqrt{3} 
    \left[1 + \frac{6.7033}{N} + O\left(N^{-2}\right) \right] \;.
\end{equation}
In Fig. \ref{plotAscaling} we give the $O\left(N^{-1}\right)$ values 
as dashed lines. Our numerical results seem to be consistent with that of 
the $1/N$ expansion.

\section{Summary and conclusions}
We have studied the CP$^{N-1}$ model on a square lattice for $N=10, 21$ 
and $41$. 
The CP$^{N-1}$ model has served as a toy model of QCD, since it shares 
fundamental properties with QCD. Here we are concerned with severe slowing
related with topological modes that plagues Monte Carlo
simulations of the two-dimensional model as well as lattice QCD.

As basic update algorithm we used a hybrid of the local heat-bath algorithm and 
the overrelaxation algorithm.  First we confirmed that standard simulations of 
systems with periodic boundary conditions in both direction suffer from a severe 
slowing down. The autocorrelation time $\tau_{int,\chi_t}$ 
of the topological susceptibility 
seems to increase exponentially with the correlation length. This problem becomes
more severe with increasing $N$. The rapid increase of $\tau_{int,\chi_t}$ means 
that for a given CPU budget there is a quite sharp bound for the correlation 
length that can be reached by simulations. In our case $\xi_{max} \approx
23, 6, 2.4$ for $N=10$, $21$, and $41$, respectively.

In order to fight the problem we followed two different strategies. 
First we tested the proposal of ref. \cite{LuSc11}. By using open 
boundary conditions in temporal direction, the problem of disconnected 
sectors in configuration space is solved by brute force.  
A nice feature of this approach is that
it should work quite independently on the Monte Carlo algorithm that is
used for the simulation.  The proposal has been tested before at the 
example of lattice QCD in ref. \cite{LuSc11,McGlMa14} and is used 
now in large scale simulations of lattice QCD.  Here we just intend to 
consolidate the findings. Simulating the two-dimensional toy model, 
high statistical accuracy can be reached and a larger range of the 
correlation length can be studied.   
In the case of $N=10$ it is hard to demonstrate a clear cut advantage for the 
open boundary conditions, since still with periodic boundary conditions a 
correlation length of a bit more than 20 can be reached.
Going to $N=21$ the situation 
becomes more clear. In this case we could obtain accurate results for the 
topological susceptibility by using open boundary conditions for values of 
$\beta$, where standard simulations with periodic boundary conditions
completely fail. While autocorrelation times are large in this range, 
slowing down is compatible with $z \approx 1$, which is typical for 
the overrelaxation algorithm. Our simulations for $N=41$  affirm the findings
for $N=21$. The results obtained here for open boundary conditions might well
serve as benchmark for new algorithmic ideas.

Furthermore we tested parallel tempering using a line defect. To this end
we introduced a sequence of systems that interpolate between a system, where 
the coupling is completely switched off along a line of length $l_d$ and 
a homogeneous system with periodic boundary conditions. For $l_d=L$ we recover
open boundary conditions. Using $l_d=L$ it turns out that a quite large 
number $N_t$ of interpolating systems is needed to get reasonable acceptance 
rates in the swaps of configurations between adjacent systems.  It turns
out that $l_d \approx \xi $ is a good choice.  We introduced an
hierarchical update scheme, where sublattices of varying size that are centred
around the defect are updated in between swaps of configurations. Focussing 
on the statistical error of the topological susceptibility, the 
parallel tempering simulation with periodic boundary 
conditions outperforms the standard simulation with open boundary conditions
by a small factor. The parallel tempering of the line defect has a number of 
free parameters. Here we fixed most of them by using simple guiding lines.
We expect that by fine tuning the performance could be improved by a small
factor. This however would require a considerable effort.  The more interesting 
question that we like to attack next is, whether the parallel tempering in a
defect structure
is helpful in lattice QCD. In particular going to dynamical fermions, the 
non-local pseudo-fermion action could be an obstacle. Here the 
combination of domain decomposition \cite{LuescherDD} with the multi-boson
algorithm \cite{LuescherMB}, as discussed in ref. \cite{Giustietal}, 
could be a solution.

Based on our simulations we obtained accurate estimates for 
$\lim_{\xi_{2nd} \rightarrow \infty} \chi_t \xi_{2nd}^2=0.01737(8)$, 
$0.00767(5)$, and $0.00391(2)$
for $N=10$, $21$, and $41$, respectively. As already pointed out in ref. 
\cite{Vicari04},
these results seem to be at odds with the $O(1/N)$ correction to the large 
$N$-limit given in ref. \cite{CaRo91}.

\section{Acknowledgement}
I thank Stefan Schaefer for discussions. 
This work was supported by the Deutsche Forschungsgemeinschaft (DFG) under the grant No HA 3150/4-1.

\end{document}